%% file: majoronLFV.tex
\documentclass[12pt]{article}
\pdfoutput=1

\usepackage{ulem}

\usepackage[utf8]{inputenc}
\usepackage[left=2.55cm, right=2.55cm, top=2.55cm, bottom=2.55cm]{geometry}
\usepackage{amsmath,amssymb,amsbsy}
\usepackage{slashed}
\usepackage{xcolor}
\usepackage{graphicx}
\usepackage{url}
\usepackage{cancel}
\usepackage{cite}
\usepackage[colorlinks=true,allcolors=darkpurple,pdfborder={0 0 0},linktocpage=false]{hyperref}
\usepackage{tabularx,booktabs}
\usepackage{multicol}
\usepackage{units}
\usepackage{xspace}
\usepackage[labelfont=bf]{caption}
\usepackage[section]{placeins}
\usepackage{subcaption}
\usepackage{soul} % para tachar
\usepackage{changepage}

\definecolor{darkred}{rgb}{0.6,0,0}
\definecolor{darkpurple}{rgb}{0.5,0,0.5}

\def\hc{\text{h.c.}}

\def\id{\mathbb{I}}
\def\z2{$\mathbb{Z}_2$}
\def\321{$\mathrm{SU(3)_c} \times \mathrm{SU(2)_L} \times \mathrm{U(1)_Y}$}
\def\one{\ensuremath{\mathbf{1}}}
\def\two{\ensuremath{\mathbf{2}}}
\def\three{\ensuremath{\mathbf{3}}}
\def\threeS{\ensuremath{\mathbf{\bar 3}}}
\definecolor{avblue}{rgb}{0.0, 0.0, 0.8}

\newcommand{\AddrIFIC}{%
  Instituto de F\'{i}sica Corpuscular, CSIC-Universitat de Val\`{e}ncia, 46980 Paterna, Spain}

\newcommand{\AddrFISTEO}{%
  Departament de F\'{\i}sica Te\`{o}rica, Universitat de Val\`{e}ncia, 46100 Burjassot, Spain}

\begin{document}

%\vspace*{-2cm}
%\begin{flushright}
%IFIC/21-XX \\
%\vspace*{2mm}
%\today
%\end{flushright}

\begin{center}
\vspace*{15mm}

\vspace{1cm}
{\Large \bf 
Observable flavor violation from spontaneous lepton number breaking
} \\
\vspace{1cm}

{\bf Pablo Escribano$^{\text{a}}$, Martin Hirsch$^{\text{a}}$, Jacopo Nava$^{\text{a}}$, Avelino Vicente$^{\text{a,b}}$}

 \vspace*{.5cm} 
 $^{(\text{a})}$ \AddrIFIC \\\vspace*{.2cm} 
 $^{(\text{b})}$ \AddrFISTEO

 \vspace*{.3cm} 
\href{mailto:pablo.escribano@ific.uv.es}{pablo.escribano@ific.uv.es}, \href{mailto:mahirsch@ific.uv.es}{mahirsch@ific.uv.es}, \href{mailto:jacopo.nava@ific.uv.es}{jacopo.nava@ific.uv.es}, \href{mailto:avelino.vicente@ific.uv.es}{avelino.vicente@ific.uv.es}
\end{center}

\vspace*{10mm}
\begin{abstract}\noindent\normalsize
  We propose a simple model of spontaneous lepton number violation
  with potentially large flavor violating decays, including the
  possibility that majoron emitting decays, such as $\mu \to e \, J$,
  saturate the experimental bounds. In this model the majoron is a
  singlet-doublet admixture. It generates a type-I seesaw for neutrino
  masses and contains also a vector-like lepton. As a by-product, the
  model can explain the anomalous $(g-2)_{\mu}$ in parts of its
  parameter space, where one expects that the branching ratio of the
  Higgs to muons is changed with respect to Standard Model
  expectations. However, the explanation of the muon $g-2$ anomaly
    would lead to tension with recent astrophysical bounds on the
    majoron coupling to muons.
\end{abstract}

\input{tex/intro}

\input{tex/model}

\input{tex/couplings}

\input{tex/pheno}

\input{tex/results}

\input{tex/conclusions}

\section*{Acknowledgements}

The authors are grateful to Isabel Cordero-Carri\'on for fruitful
discussions. Work supported by the Spanish grants FPA2017-85216-P
(MINECO/AEI/FEDER, UE) and SEJI/2018/033 and PROMETEO/2018/165
(Generalitat Valenciana). The work of PE is supported by the FPI grant
PRE2018-084599. AV acknowledges financial support from MINECO through
the Ramón y Cajal contract RYC2018-025795-I.

\appendix

\input{tex/app1}

\input{tex/app2}

\input{tex/app3}

\bibliographystyle{utphys}
\bibliography{refs}

\end{document}

%% file: tex/intro.tex
\section{Introduction}
\label{sec:intro}

Neutrino masses may be non-zero due to the violation of lepton number,
$L$, in which case neutrinos are Majorana particles. The literature is
abound with Majorana neutrino mass models which simply add explicit
lepton number violating interactions or mass terms to the Lagrangian,
but the violation of $L$ could well be spontaneous in origin. The
spontaneous breaking of a global continuous quantum number generates a
massless Goldstone boson, in the case of lepton number usually called
the \textit{majoron}~\cite{Chikashige:1980qk,Chikashige:1980ui}, $J$.
The original majoron of~\cite{Chikashige:1980qk,Chikashige:1980ui} is
a complete gauge singlet. On the other hand, it is also possible to
use larger multiplets to break lepton number spontaneously, resulting
in the doublet and triplet
majorons~\cite{Schechter:1981cv,Gelmini:1980re,Aulakh:1982yn}. In
general, the majoron can be an admixture of these three
representations, or even more exotic possibilities.

A massless boson in the particle spectrum certainly will affect
phenomenology.  However, whether these changes with respect to the
explicit lepton number violating models are quantitatively important
depends very strongly on the model and the nature of the majoron. The
pure singlet majoron interacts so weakly with all Standard Model (SM)
particles, that it is highly unlikely it will ever be observed
experimentally. In the other extreme, \textit{pure} doublet and
triplet majorons have been ruled out by
LEP~\cite{ParticleDataGroup:2020ssz}, but majorons with sufficiently
large singlet admixture can escape this constraint.

The interaction of the majoron with charged leptons can be described
in a model independent way as~\cite{Escribano:2020wua},
\begin{equation} \label{eq:llJ}
\mathcal{L}_{\ell \ell J} = J \, \bar{\ell}_\beta \left( S_L^{\beta \alpha} \, P_L + S_R^{\beta \alpha} \, P_R \right)\ell_{\alpha} + \hc \, ,
\end{equation}
where $\ell_{\alpha,\beta}$ are the standard light charged leptons and
$P_{L,R}$ are the usual chiral projectors. The $S_{L,R}$ couplings are
dimensionless coefficients, and we consider all flavor combinations:
$\beta\alpha = \left\{
ee,\mu\mu,\tau\tau,e\mu,e\tau,\mu\tau\right\}$. Due to the
pseudoscalar nature of majorons, the diagonal $S^{\beta \beta} =
S_L^{\beta \beta} + S_R^{\beta \beta \ast}$ couplings are purely
imaginary.

Apart from the invisible $Z$-boson width, measured at LEP, there are a
number of laboratory and astrophysical constraints on majorons. First,
neutrinoless double beta decay will occur with the emission of a
majoron. The best current constraints come from the EXO-200
experiment~\cite{EXO-200:2014vam}, limiting the effective coupling of
the majoron to electron-type neutrinos, to values roughly below ${\cal
  O}(10^{-5})$. Astrophysics also constrains the majoron parameter
space. The production of majorons inside stars and their posterior
emission constitutes a very efficient stellar cooling mechanism. This
allows one to set very stringent constraints on the majoron couplings
to charged leptons. For instance, white dwarfs allowed the authors
of~\cite{Calibbi:2020jvd} to set the bound $\text{Im} \, S^{e e} < 2.1
\times 10^{-13}$, whereas Ref.~\cite{Croon:2020lrf} considered the
supernova SN1987A and found $\text{Im} \, S^{\mu \mu} < 2.1 \times
10^{-10}$. Last but not least, one can also derive indirect bounds on
the majoron couplings to charged leptons from the bounds on the
majoron couplings to photons, since the former induce the latter at
the 1-loop level. Using results from the OSQAR
experiment~\cite{Ballou:2014myz}, a light-shining-through-a-wall
experiment, Ref.~\cite{Escribano:2020wua} found the approximate bounds
$S^{ee} \lesssim 10^{-7}$ and $S^{\mu\mu} \lesssim 10^{-5}$. However,
these are less stringent than the stellar cooling bounds mentioned
above.

In this paper we study charged lepton flavor violation (LFV),
connected to the spontaneous violation of lepton number. While our
numerical results are obtained for one particular model realization,
it is possible to discuss some general features qualitatively. The
$S_{L,R}$ couplings in Eq.~\eqref{eq:llJ} can be generically written
as
\begin{equation}\label{eq:SLR}
  S_{L,R} \sim a \, \frac{\widehat M_\ell}{v} + b \, Y \, .
\end{equation}
Here, $Y$ is a general matrix in flavor space and ${\widehat M_\ell}$
is the diagonal charged lepton mass matrix, $\widehat M_\ell =
\text{diag}(m_e,m_\mu,m_\tau)$. The coefficients $a$ and $b$ depend on
the model under consideration. In the case of a type-I seesaw and a
pure singlet (or doublet) majoron, both $a$ and $b$ are zero at
tree-level, but generated at
1-loop~\cite{Chikashige:1980ui,Pilaftsis:1993af,Heeck:2019guh}. The
1-loop diagram is further suppressed by the small mixing between the
right-handed neutrinos and the active states, thus one expects that
decays such as $\mu \to e \, J$ are unobservable. For the triplet
majoron, on the other hand, $a$ is non-zero at tree-level. It is
generated via the mixing of the triplet with the SM Higgs, due to a
coupling of the form $\lambda \, \sigma H \Delta H$, where $\sigma$ is
the scalar singlet, $H$ the SM Higgs and $\Delta$ the triplet
scalar. $b$ again can be generated only radiatively and LFV
interactions with majorons are expected to be tiny.  How about
type-III seesaw? In the type-III seesaw, the charged leptons of the SM
are mixed with the charged components of the fermionic triplet
$\Sigma$. In the spontaneous version of this setup one would thus
expect some non-diagonal coupling of the majoron to charged leptons to
appear in the mass eigenstate basis~\cite{Cheng:2020rla}. However, the
corresponding mixing is related to the small neutrino masses and thus,
in the end, the rates for $\ell_\alpha \to \ell_\beta \, J$ decays are
typically very small.

This discussion can serve as a basis to establish the criteria a model
has to fulfill in order to have sizeable off-diagonal couplings
between the majoron and charged leptons. First of all, if the majoron
couplings to charged leptons are induced via majoron mixing with the
SM Higgs doublet, they will be exactly diagonal in the charged lepton
mass basis. Therefore, in order to obtain sizable off-diagonal
couplings, the majoron must couple directly to the charged lepton
sector. This coupling can be either to the light charged leptons
themselves or to some additional heavy charged leptons which, after
symmetry breaking, mix with them. In the first case, sizeable
off-diagonal couplings can be obtained with a non-universal lepton
number assignment, whereas in the second case it is crucial that the
light-heavy mixing is not suppressed by neutrino masses. We refer
to~\cite{Sun:2021jpw} for a discussion along similar lines.

In this paper we propose a relatively simple model that induces large
off-diagonal majoron couplings to charged leptons at tree-level. Our
model adds a singlet and a doublet scalar to the SM, both with lepton
number. It also extends the SM symmetry by imposing lepton number
conservation. Finally, we introduce three right-handed neutrinos and a
vector-like lepton, which we choose to be an $\rm SU(2)_L$ singlet for
simplicity. After the electroweak and lepton number symmetries are
spontaneously broken, neutrinos acquire non-zero masses via a
TeV-scale type-I seesaw mechanism and the vector-like lepton mixes
with the SM charged leptons, inducing in this way large LFV majoron
couplings. Furthermore, extending the SM lepton sector with
vector-like fermions will affect a number of observables, most notably
the anomalous magnetic moment of the charged leptons, as well as their
coupling to gauge bosons. As a by-product of our construction, the
model can explain the observed anomaly in the muon anomalous magnetic
moment~\cite{Aoyama:2020ynm,Muong-2:2021ojo} in parts of its parameter
space. It can also lead to observable effects in Higgs boson decays,
most notably in $h \to \mu \mu$.

We mention in passing that branching ratios for $\mu \to e \, J$
decays as large at the experimental limit have been found
in~\cite{Hirsch:2009ee}.  The underlying model is supersymmetric with
spontaneous violation of R-parity. This provides one, albeit quite
complicated, model example, where off-diagonal majoron couplings
to charged leptons are induced at tree-level and can be
large. Nevertheless, we note that this decay can in principle
  saturate the experimental bound also in models that generate the
  majoron couplings to charged leptons at the 1-loop
  level~\cite{Heeck:2019guh}.

The rest of this paper is organized as follows. In the next Section we
introduce the model, discuss the majoron profile and the mass matrices
for leptons and scalars. Section~\ref{sec:coups} is devoted to a
discussion of the majoron couplings, while Section~\ref{sec:pheno}
discusses the possible phenomenological signatures. In
Section~\ref{sec:results} we present our numerical results, before
closing the paper with a short summary in
Section~\ref{sec:conclusions}. Some more technical aspects of our
calculations are given in the Appendices.

%% file: tex/model.tex
\section{The model}
\label{sec:model}

We consider a type-I seesaw with spontaneous lepton number
violation. The quark sector remains as in the SM, whereas the lepton
sector is extended with the addition of $3$ generations of singlet
right-handed neutrinos, a pair of singlet vector-like leptons, $F_L$
and $F_R$, the scalars $\sigma$ and $S$ and a $\rm U(1)_L$ global
symmetry, where $\rm L$ refers to lepton number. The full particle
content of the model and the representations of all fields under the
gauge and global groups are shown in Tab.~\ref{tab:content}.

As usual, the SM $\rm SU(2)_L$ doublets can be decomposed as
\begin{equation}
q_L = \left( \begin{array}{c}
u \\
d
\end{array} \right)_L \quad , \quad \ell_L = \left( \begin{array}{c}
\nu \\
e
\end{array} \right)_L \quad , \quad H = \left( \begin{array}{c}
H^+ \\
H^0
\end{array} \right) \, ,
\end{equation}
whereas the new $S$ doublet can be decomposed as
\begin{equation}
S = \left( \begin{array}{c}
S^0 \\
S^-
\end{array} \right) \, .
\end{equation}
%%%%%%%%%%%%%%%%%%%%PARTICLE CONTENT%%%%%%%%%%%%%%%%%%%%%%%%%%%%%%%%%%%%%%%%
{
\renewcommand{\arraystretch}{1.6}
\begin{table}[t]
\centering
\begin{tabular}{ c | c c c c c c c | c c c }
\toprule
& $q_L$ & $u_R$ & $d_R$ & $\ell_L$ & $e_R$ & $\nu_R$ & $F_{L,R}$ & $H$ & $\sigma$ & $S$ \\ 
\hline
$\rm SU(3)_C$ & $\three$ & $\threeS$ & $\threeS$ & $\one$ & $\one$ & $\one$ & $\one$ & $\one$ & $\one$ & $\one$ \\
$\rm SU(2)_L$ & $\two$ & $\one$ & $\one$ & $\two$ & $\one$ & $\one$ & $\one$ & $\two$ & $\one$ & $\two$ \\
$\rm U(1)_Y$ & $\frac{1}{6}$ & $\frac{2}{3}$ & $-\frac{1}{3}$ & $-\frac{1}{2}$ & $-1$ & $0$ & $-1$ & $\frac{1}{2}$ & $0$ & $-\frac{1}{2}$ \\[1mm]
\hline
$\rm U(1)_L$ & $0$ & $0$ & $0$ & $1$ & $1$ & $1$ & $-1$ & $0$ & $2$ & $-2$ \\
%\hline
\textsc{Generations} & 3 & 3 & 3 & 3 & 3 & 3 & 1 & 1 & 1 & 1 \\
\bottomrule
\end{tabular}
\caption{Particle content of the model and representations under the
  gauge and global symmetries. $q_L$, $\ell_L$, $u_R$, $d_R$, $e_R$
  and $H$ are the usual SM fields.}
\label{tab:content}
\end{table}
}
%%%%%%%%%%%%%%%%%%%%%%%%%%%%%%%%%%%%%%%%%%%%%%%%%%%%%%%%%%%%%%%%%%%%%%%%%%%%%%
Under the above working assumptions, the most general Yukawa
Lagrangian allowed by all symmetries can be written as
\begin{equation} \label{eq:lag}
\mathcal{L}_Y = \mathcal{L}_Y^{\text{SM}} + \mathcal{L}_Y^{\text{extra}} \, ,
\end{equation}
where
\begin{equation} \label{eq:YukawaSM}
-\mathcal{L}_Y^{\text{SM}} =
  Y_u \, \overline{u}_R \, q_L \, H
  + Y_d \, \overline{d}_R \, q_L \,\widetilde{H}
  + Y_e \, \overline{e}_R \, \ell_L \,\widetilde{H}
  + \hc 
\end{equation}
is the usual SM Lagrangian, with $\widetilde{H}=i\,\tau_2\,H^*$ and
$Y_{u,d,e}$ $3\times 3$ Yukawa matrices in flavor space. We have
omitted $\rm SU(2)_L$ contractions and flavor indices to simplify
the notation. The new terms are given by
\begin{align} 
  -\mathcal{L}_Y^{\text{extra}} =&
  Y_\nu \, \overline{\nu}_R \, \ell_L \, H
  + \frac{1}{2} \, \kappa \, \sigma \, \overline{\nu_R} \, \nu_R^c
  + \rho \, \sigma \, \overline{e}_R \, F_L
  + Y_S \, S \, \overline{F}_R \, \ell_L
  + M_F \, \overline{F}_R \, F_L
  + \hc \, . \label{eq:Yextra}
\end{align}
Here $Y_\nu$ and $\kappa$ are $3\times 3$ matrices, $\rho$ is a $3
\times 1$ matrix and $Y_S$ is $1 \times 3$ matrix. The parameter $M_F$
has dimensions of mass and, again, $\rm SU(2)_L$ contractions and
flavor indices have been omitted for the sake of clarity. The guiding
principle when writing Eq.~\eqref{eq:Yextra} is the conservation of
lepton number.~\footnote{For instance, we have not included a Yukawa
  term of the form $\sigma \, \overline{e}_R \, F_R^c$ because it
  would violate lepton number explicitly.} Finally, the scalar
potential of the model also includes new terms involving the $\sigma$
and $S$ fields. It can be written as
\begin{equation} \label{eq:pot}
  \mathcal V = \mathcal V_H + \mathcal V_\sigma + \mathcal V_S + \mathcal V_{\rm mix} \, ,
\end{equation}
where
\begin{equation}
  \mathcal V_\phi = m_\phi^2 |\phi|^2 + \frac{\lambda_\phi}{2} \, |\phi|^4 \, ,
\end{equation}
with $\phi = H, \sigma, S$, and
\begin{equation}
  \mathcal V_{\rm mix} =
  \lambda_{H \sigma} \, |H|^2 \, |\sigma|^2
  + \lambda_{H S}^{(1)} \, |H|^2 \, |S|^2
  + \lambda_{H S}^{(2)} \, H^\dagger \, S \, S^\dagger \, H
  + \lambda_{\sigma S} \, |\sigma|^2 \, |S|^2
  + \left( \mu \, H \, \sigma \, S + \hc \right) \, . \label{eq:potmix}
\end{equation}
Here all $m_\phi^2$ parameters have dimensions of mass$^2$, whereas
$\mu$ is a parameter with dimensions of mass.

\subsection{Scalar sector}
\label{subsec:scalar}

The scalars of the model take the vacuum expectation values (VEVs)
\begin{equation}
  \langle H \rangle = \frac{1}{\sqrt{2}} \, \left( \begin{array}{c}
    0 \\
    v_H
  \end{array} \right) \, , \quad
  \langle \sigma \rangle = \frac{v_\sigma}{\sqrt{2}} \, , \quad
  \langle S \rangle = \frac{1}{\sqrt{2}} \, \left( \begin{array}{c}
    v_S \\
    0
\end{array} \right) \, .
\end{equation}
These relations define the VEVs $v_H$, $v_\sigma$ and $v_S$, which
break the electroweak and $\rm U(1)_L$ symmetries. As a result of
this, the $W$ and $Z$ gauge bosons acquire non-zero masses, given by
\begin{align}
  m_W^2 &= \frac{1}{4} \, g^2 \, v^2 \, , \\
  m_Z^2 &= \frac{1}{4} \, \left(g^2+{g^\prime}^2\right) \, v^2\, ,
\end{align}
where $v^2 = v_H^2 + v_S^2$ and $g$ and $g^\prime$ are the $\rm
SU(2)_L$ and $\rm U(1)_Y$ gauge couplings, respectively. Here $v
\simeq 246$ GeV is the usual electroweak VEV, which receives
contributions from both scalar doublets $H$ and $S$. The tadpole
equations obtained by minimizing the scalar potential read
\begin{align}
    \frac{\partial \mathcal{V}}{\partial H^0} &= \frac{v_H}{\sqrt{2}} \left( m_H^2 + \frac{v_H^2 \, \lambda_H}{2} + \frac{v_S^2 \, \lambda_{HS}^{(1)}}{2} + \frac{v_\sigma^2 \, \lambda_{H\sigma}}{2} - \frac{v_S \, v_\sigma \, \mu}{\sqrt{2} \, v_H} \right) = 0 \, , \label{eq:min_pot_1} \\
    \frac{\partial \mathcal{V}}{\partial \sigma} &= \frac{v_\sigma}{\sqrt{2}} \left( m_\sigma^2 + \frac{v_\sigma^2 \, \lambda_\sigma}{2} + \frac{v_H^2 \, \lambda_{H\sigma}}{2} + \frac{v_S^2 \, \lambda_{S\sigma}}{2} - \frac{v_H \, v_S \, \mu}{\sqrt{2} \, v_\sigma} \right) = 0 \, , \label{eq:min_pot_2} \\
    \frac{\partial \mathcal{V}}{\partial S^0} &= \frac{v_S}{\sqrt{2}} \left( m_S^2 + \frac{v_S^2 \,\lambda_S}{2} + \frac{v_H^2 \, \lambda_{HS}^{(1)}}{2} + \frac{v_\sigma^2 \, \lambda_{S\sigma}}{2} - \frac{v_H \, v_\sigma \, \mu}{\sqrt{2} \, v_S} \right) = 0 \, . \label{eq:min_pot_3}
\end{align}
The trilinear $\mu$ term in Eq.~\eqref{eq:potmix} plays an important
role. In the limit $\mu \to 0$, the scalar potential has an accidental
$\rm U(1)$ symmetry, under which all scalar fields can have arbitrary
charges. Therefore, its presence is crucial to explicitly break this
global symmetry and avoid the appearance of an unwanted Goldstone
boson. The $\mu$ term also induces a tadpole for each of the scalar
fields if the other two have non-vanishing VEVs.

Assuming that CP is not violated in the scalar sector, namely that all
scalar potential parameters and VEVs are real, we can split the
neutral scalar fields into their real and imaginary components as
\begin{align}
\label{eq:neutH} H^0 &= \frac{1}{\sqrt{2}} \left( S_H + i \, P_H + v_H \right) \, , \\
\label{eq:neutsigma} \sigma &= \frac{1}{\sqrt{2}} \left( S_\sigma + i \, P_\sigma + v_\sigma \right) \, , \\
\label{eq:neutS} S^0 &= \frac{1}{\sqrt{2}} \left( S_S + i \, P_S + v_S \right) \, .
\end{align}
The scalar potential contains the piece $\mathcal V_{\rm mass} =
\mathcal V_{\rm mass}^N + \mathcal V_{\rm mass}^C$, with mass terms
for the neutral ($\mathcal V_{\rm mass}^N$) and charged ($\mathcal
V_{\rm mass}^C$) scalars in the model. The neutral scalars mass terms
read
\begin{equation}
  \mathcal V_{\rm mass}^N = \frac{1}{2} \, \text{Re}(z_i) \, \left( \mathcal{M}_R^2 \right)_{ij} \, \text{Re}(z_j) + \frac{1}{2} \, \text{Im}(z_i) \, \left( \mathcal{M}_I^2 \right)_{ij} \, \text{Im}(z_j) \, ,
\end{equation}
where $z = \{H^0, \sigma, S^0 \}$ and $\mathcal{M}_R^2$ and
$\mathcal{M}_I^2$ are the $3 \times 3$ squared mass matrices for the
CP-even and CP-odd neutral states, respectively. The prefactors of
$\frac{1}{2}$ are due to the fact that $\text{Re}(z_i)$ and
$\text{Im}(z_i)$ are real scalar fields. It is straightforward to get
the analytical expressions of the mass matrices, which can be computed
as
\begin{align}
  \left( \mathcal{M}_R^2 \right)_{ij} &= \frac{1}{2} \left( \frac{\partial^2 \mathcal V_{\rm mass}}{\partial z_i \partial z_j} + \frac{\partial^2 \mathcal V_{\rm mass}}{\partial z_i^\ast \partial z_j^\ast} \right) + \frac{\partial^2 \mathcal V_{\rm mass}}{\partial z_i \partial z_j^\ast} \, , \\
  \left( \mathcal{M}_I^2 \right)_{ij} &= - \frac{1}{2} \left( \frac{\partial^2 \mathcal V_{\rm mass}}{\partial z_i \partial z_j} + \frac{\partial^2 \mathcal V_{\rm mass}}{\partial z_i^\ast \partial z_j^\ast} \right) + \frac{\partial^2 \mathcal V_{\rm mass}}{\partial z_i \partial z_j^\ast} \, .
\end{align}
Then, using the expressions above, one finds~\footnote{It proves
  useful to compute the matrix $\mathcal{M}_I^2$ in a general $R_\xi$
  gauge, since this allows for the proper identification of the
  Goldstone boson that becomes the longitudinal component of the $Z$
  boson. However, we present here the results in Landau gauge ($\xi =
  0$).}
\begin{equation} \label{eq:MR2}
  \hspace{-1.2cm} \mathcal{M}_R^2 = \left( \begin{array}{ccc}
    m_H^2 + \frac{3\,v_H^2\,\lambda_H}{2} + \frac{v_S^2\,\lambda_{HS}^{(1)}}{2} + \frac{v_\sigma^2\,\lambda_{H\sigma}}{2} & v_H \, v_\sigma \, \lambda_{H\sigma} - \frac{v_S \, \mu}{\sqrt{2}} & v_H \, v_S \, \lambda_{HS}^{(1)} - \frac{v_\sigma \, \mu}{\sqrt{2}} \\
    v_H \, v_\sigma \, \lambda_{H\sigma} - \frac{v_S \, \mu}{\sqrt{2}} & m_\sigma^2 + \frac{3\,v_\sigma^2\,\lambda_\sigma}{2} + \frac{v_H^2\,\lambda_{H\sigma}}{2} + \frac{v_S^2\,\lambda_{S\sigma}}{2} & v_S \, v_\sigma \, \lambda_{S\sigma} - \frac{v_H \, \mu}{\sqrt{2}} \\
    v_H \, v_S \, \lambda_{HS}^{(1)} - \frac{v_\sigma \, \mu}{\sqrt{2}} & v_S \, v_\sigma \, \lambda_{S\sigma} - \frac{v_H \, \mu}{\sqrt{2}} & m_S^2 + \frac{3\,v_S^2\,\lambda_S}{2} + \frac{v_H^2\,\lambda_{HS}^{(1)}}{2} + \frac{v_\sigma^2\,\lambda_{S\sigma}}{2}
    \end{array} \right) \, ,
\end{equation}
and
\begin{equation} \label{eq:MI2}
  \hspace{-0.9cm} \mathcal{M}_I^2 = \left( \begin{array}{ccc}
    m_H^2 + \frac{v_H^2\,\lambda_H}{2} + \frac{v_S^2\,\lambda_{HS}^{(1)}}{2} + \frac{v_\sigma^2\,\lambda_{H\sigma}}{2} & \frac{v_S \, \mu}{\sqrt{2}} & \frac{v_\sigma \, \mu}{\sqrt{2}} \\
    \frac{v_S \, \mu}{\sqrt{2}} & m_\sigma^2 + \frac{v_\sigma^2\,\lambda_\sigma}{2} + \frac{v_H^2\,\lambda_{H\sigma}}{2} + \frac{v_S^2\,\lambda_{S\sigma}}{2} & \frac{v_H\, \mu}{\sqrt{2}} \\
    \frac{v_\sigma \, \mu}{\sqrt{2}} & \frac{v_H \, \mu}{\sqrt{2}} & m_S^2 + \frac{v_S^2\,\lambda_S}{2} + \frac{v_H^2\,\lambda_{HS}^{(1)}}{2} + \frac{v_\sigma^2\,\lambda_{S\sigma}}{2}
    \end{array} \right) \, .
\end{equation}
One can now use the tadpole equations in
Eqs.~\eqref{eq:min_pot_1}-\eqref{eq:min_pot_3} to evaluate these
matrices at the minimum of the scalar potential. We obtain
\begin{equation} \label{eq:MR2b}
  \mathcal{M}_R^2 = \left( \begin{array}{ccc}
    v_H^2 \, \lambda_H + \frac{v_S \, v_\sigma \, \mu}{\sqrt{2} \, v_H} & v_H \, v_\sigma \, \lambda_{H\sigma} - \frac{v_S \, \mu}{\sqrt{2}} & v_H \, v_S \, \lambda_{HS}^{(1)} - \frac{v_\sigma \, \mu}{\sqrt{2}} \\
    v_H \, v_\sigma \, \lambda_{H\sigma} - \frac{v_S \, \mu}{\sqrt{2}} & v_\sigma^2 \, \lambda_\sigma + \frac{v_H \, v_S \, \mu}{\sqrt{2} \, v_\sigma} & v_S \, v_\sigma \, \lambda_{S\sigma} - \frac{v_H \, \mu}{\sqrt{2}} \\
    v_H \, v_S \, \lambda_{HS}^{(1)} - \frac{v_\sigma \, \mu}{\sqrt{2}} & v_S \, v_\sigma \, \lambda_{S\sigma} - \frac{v_H \, \mu}{\sqrt{2}} & v_S^2 \, \lambda_S + \frac{v_H \, v_\sigma \, \mu}{\sqrt{2} \, v_S}
    \end{array} \right) \, ,
\end{equation}
and
\begin{equation} \label{eq:MI2b}
  \mathcal{M}_I^2 = \left( \begin{array}{ccc}
    \frac{v_S \, v_\sigma \, \mu}{\sqrt{2} \, v_H} & \frac{v_S \, \mu}{\sqrt{2}} & \frac{v_\sigma \, \mu}{\sqrt{2}} \\
    \frac{v_S \, \mu}{\sqrt{2}} & \frac{v_H \, v_S \, \mu}{\sqrt{2} \, v_\sigma} & \frac{v_H \, \mu}{\sqrt{2}} \\
    \frac{v_\sigma \, \mu}{\sqrt{2}} & \frac{v_H \, \mu}{\sqrt{2}} & \frac{v_H \, v_\sigma \, \mu}{\sqrt{2} \, v_S}
    \end{array} \right) \, .
\end{equation}

The physical CP-even mass eigenstates $\{H_1, H_2, H_3\}$  are related to the corresponding weak eigenstates $\{S_H, S_{\sigma}, S_S\}$  as
\begin{equation} \label{eq:CPeven}
 \left( \begin{array}{c}
    H_1\\
    H_2\\
    H_3 \end{array} \right) = W \, \left( \begin{array}{c}
    S_H\\
    S_{\sigma}\\
    S_S \end{array} \right) \, ,
\end{equation}
where $W$ is the $3\times 3$ unitary matrix which brings the
matrix $\mathcal{M}_R^2$ into diagonal form as
\begin{equation}
W \, \mathcal{M}_R^2 \, W^T = \text{diag} (m_{H_1}^2, m_{H_2}^2, m_{H_3}^2) \, .
\end{equation}
The model has thus $3$ CP-even neutral scalars. One of them,
presumably the lighest, is to be identified with the Higgs boson
discovered at the LHC, $H_1$, with $m_{H_1} \approx 125$
GeV. Similarly, diagonalizing the mass matrix $\mathcal{M}_I^2$, we
can obtain the profile of the three CP-odd mass eigenstates. We end up
with a massive state, that we denote by $A$, and two massless
states. One of the massless states is the Goldstone boson, $z$,
\textit{eaten} by the $Z$ boson, while the other is the majoron, $J$,
the Goldstone boson associated to the spontaneous breaking of lepton
number. In the basis $\text{Im} \, \{H^0, \sigma, S^0 \} = \{ P_H,
P_\sigma, P_S \}$, the mass eigenstates are given in terms of the
original gauge eigenstates as
\begin{align}
  z &= \frac{1}{v} \, \left( v_H , 0 , -v_S \right) \, , \\
  J &= \frac{1}{V^2} \, \left( \frac{v_H \, v_S^2}{v} , - v \, v_\sigma , \frac{v_H^2 \, v_S}{v} \right) \, , \label{eq:J} \\
  A &= \frac{1}{V^2} \, \left( v_S \, v_\sigma , v_H \, v_S , v_H \, v_\sigma \right) \, , \label{eq:A}
\end{align}
with their masses given by
\begin{equation} \label{eq:mA}
  m_z^2 = m_J^2 = 0 \, , \quad m_A^2 = \frac{\mu \, V^4}{\sqrt{2} \, v_H \, v_S \, v_\sigma} \, .
\end{equation}
We have defined the combination $V^4 = v_H^2 \, v_S^2 + v_H^2 \,
v_\sigma^2 + v_S^2 \, v_\sigma^2$. As already discussed, the $\mu$
parameter breaks an accidental $\rm U(1)$ symmetry that would lead in
its absence to the appearance of an additional massless Goldstone
boson. This can be observed in $m_A^2$, that would vanish if $\mu =
0$. We have also found that the majoron has a non-vanishing component
in the doublet directions. Therefore, in order to avoid
phenomenological problems with a doublet majoron, such as a sizable
invisible $Z$-boson width, we are forced to impose the hierarchy of
VEVs
\begin{equation} \label{eq:hier}
  v_H , v_S \ll v_\sigma \, ,
\end{equation}
which guarantees that the majoron is mostly singlet. We turn now to
the charged scalar mass matrix. In this case, the scalar potential
contains the term
\begin{equation}
  \mathcal V_{\rm mass}^C = \left( \begin{array}{c c}
	H^- & S^- 
  \end{array} \right) \, \mathcal{M}_\pm^2 \left( \begin{array}{c}
	H^+ \\
	S^+
  \end{array} \right) \, ,
\end{equation}
with
\begin{equation}
  \mathcal{M}_\pm^2 = \left( \begin{array}{cc}
	m_H^2 + \frac{v_H^2 \, \lambda_{H}}{2} + \frac{v_S^2}{2} \left( \lambda_{HS}^{(1)} + \lambda_{HS}^{(2)} \right) + \frac{v_\sigma^2 \, \lambda_{H\sigma}}{2} & \frac{v_\sigma \, \mu}{\sqrt{2}} + \frac{\lambda_{HS}^{(2)} \, v_H \, v_S}{2} \\
	\frac{v_\sigma \, \mu}{\sqrt{2}} + \frac{\lambda_{HS}^{(2)} \, v_H \, v_S}{2} & m_S^2 + \frac{v_S^2 \, \lambda_{S}}{2} + \frac{v_H^2}{2} \left( \lambda_{HS}^{(1)} + \lambda_{HS}^{(2)} \right) + \frac{v_\sigma^2 \, \lambda_{S\sigma}}{2}
  \end{array} \right) \, .
\end{equation}
Again, the application of the tadpole equations leads to
\begin{equation}
  \mathcal{M}_\pm^2 = \left( \begin{array}{cc}
	\frac{v_S \, v_\sigma \, \mu}{\sqrt{2} \, v_H} + \frac{\lambda_{HS}^{(2)} \, v_S^2}{2} & \frac{v_\sigma \, \mu}{\sqrt{2}} + \frac{\lambda_{HS}^{(2)} \, v_H \, v_S}{2} \\
	\frac{v_\sigma \, \mu}{\sqrt{2}} + \frac{\lambda_{HS}^{(2)} \, v_H \, v_S}{2} & \frac{v_H \, v_\sigma \, \mu}{\sqrt{2} \, v_S} + \frac{\lambda_{HS}^{(2)} \, v_H^2}{2}
  \end{array} \right) \, .
\end{equation}
One of the eigenvalues of this matrix vanishes. This corresponds to
the Goldstone boson eaten by the $W$ boson, $w$. The other state is
the massive charged scalar $C^\pm$. In the basis $\{ H^\pm , S^\pm
\}$, they are given in terms of the gauge eigenstates as
\begin{align}
  w^\pm &= \frac{1}{v} \, \left( -v_H , v_S \right) \, , \\
  C^\pm &= \frac{1}{v} \, \left( v_S , v_H \right) \, ,
\end{align}
and their masses are
\begin{equation}
  m_w^2 = 0 \, , \quad m_C^2 = \frac{1}{2} \, \frac{v^2}{v_H \, v_S} \, \left( \sqrt{2} \, \mu \, v_\sigma + \lambda_{HS}^{(2)} \, v_H \, v_S \right)  \, .
\end{equation}

\subsection{Lepton masses}
\label{subsec:leptonmass}

The light neutrinos get their masses by means of a standard type-I
seesaw mechanism. Defining the $3 \times 3$ matrices in generation
space $m_D$ and $M_R$
\begin{equation}\label{eq:Neutrino}
  m_D = \frac{v_H}{\sqrt{2}} \, Y_\nu \, , \quad M_R = \frac{v_\sigma}{\sqrt{2}} \, \kappa \, ,
\end{equation}
the neutral leptons mass term is given by
\begin{equation}\label{eq:Neutral}
  -\mathcal{L}_N = \frac{1}{2} \left( \begin{array}{cc}
    \bar{\nu}_L^{c} & \bar{\nu}_R
  \end{array} \right) \mathcal{M}_N \left( \begin{array}{c}
    \nu_L \\
    \nu_R^{c} \end{array} \right) + \hc
\end{equation}
with the $6 \times 6$ matrix $\mathcal{M}_N$ defined as
\begin{equation}
\mathcal{M}_N = \left( \begin{array}{cc}   
    0 & m_D^T \\
    m_D & M_R \end{array} \right) \, . \end{equation}
The resulting mass matrix corresponds to the standard type-I seesaw
matrix. If $m_D \ll M_R$, the light neutrinos mass matrix is given by
the well-known formula $m_\nu = - m_D^T \, M_R^{-1} \, m_D$. We note
that the hierarchy $m_D \ll M_R$ follows naturally from the hierarchy
in Eq.~\eqref{eq:hier}. On the other hand, the mass term of the
charged leptons reads
\begin{equation}\label{eq:Charged}
  -\mathcal{L}_C = \left( \begin{array}{cc}
  \bar{e}_R & \bar{F}_R
   \end{array} \right)  \mathcal{M}_C \, \left( \begin{array}{c}
    e_L \\
    F_L \end{array} \right) + \hc \, ,
\end{equation}
where the $4 \times 4$ matrix $\mathcal{M}_C$ is given by
\begin{equation}
  \mathcal{M}_C = \left( \begin{array}{cc}   
    m_e & m_\rho \\
    m_S & M_F \end{array} \right) \, ,
\end{equation}
and we have defined
\begin{equation}
  m_e = \frac{v_H}{\sqrt{2}} \, Y_e \, , \quad m_\rho = \frac{v_\sigma}{\sqrt{2}} \, \rho \, , \quad m_S = \frac{v_S}{\sqrt{2}} \, Y_S \, .
\end{equation}
The matrices $m_e$, $m_\rho$ and $m_S$ are $3 \times 3$, $3 \times 1$
and $1 \times 3$, respectively. The mass matrices $\mathcal{M}_N$ and
$\mathcal{M}_C$ can be brought to diagonal form as
\begin{align}
  \widehat{\mathcal{M}}_N &= \text{diag} (m_{N_i}) = V^{\nu} \, \mathcal{M}_N \, V^{\nu T} \, , \label{eq:MNdiag} \\
  \widehat{\mathcal{M}}_C &= \text{diag} (m_{C_i}) = V^{R^{\dagger}} \, \mathcal{M}_C \, V^{L} \, , \label{eq:MCdiag}
\end{align}
where $V^{\nu}$ and $V^{L,R}$ are unitary matrices.~\footnote{We
  observe that for $v_H \ll v_\sigma$, which is required by the seesaw
  mechanism, and assuming all the Yukawa couplings to be of
  $\mathcal{O}(1)$, we are forced to impose $v_S \ll v_\sigma$ in
  order not to have all charged lepton masses pushed towards the
  seesaw scale. This is in agreement with the considerations regarding
  the majoron profile.}

It proves convenient to derive approximate expressions for the
matrices involved in Eq.~\eqref{eq:MCdiag}. We now
follow~\cite{Grimus:2000vj} to obtain approximate expressions for the
matrices $V^{L,R}$ by performing a perturbative expansion in powers of
the inverse of the largest scale in $\mathcal{M}_C$.~\footnote{See
also the pioneer work~\cite{Schechter:1981cv} for an alternative (but
equivalent) approach for the perturbative diagonalization of a
Majorana mass matrix.} We assume
\begin{equation} \label{eq:hierV}
  m_e , m_S \ll m_\rho \ll M_F \, ,
\end{equation}
consistent with Eq.~\eqref{eq:hier}. Then, we can write the matrices
$V^{L,R}$ as
\begin{equation}
  V^{L(R)} = U^{L(R)}\, D^{L(R)} \, ,
\end{equation}
where $U^{L,R}$ and $D^{L,R}$ are unitary matrices. The matrices
$U^{L,R}$ will be responsible for the block-diagonalization of
$\mathcal{M}_C$ while $D^{L,R}$ will diagonalize the light and heavy
sub-blocks. The matrices $D^{L,R}$ can be written in the form
\begin{equation}
   D^{L(R)}= \left( \begin{array}{cc}   
   D^{L(R)}_e & 0\\
    0 &   D^{L(R)}_F\end{array}\right) \, ,
\end{equation}
where $D^{L,R}_e$ are $3 \times 3$ matrices and $D^{L,R}_F$ are just
complex phases. While the non-vanishing elements of the $D$ matrices
are expected to be of $\mathcal{O}(1)$, the elements of $U$ will
instead be sensitive to the hierarchy of the scales involved in
$\mathcal{M}_C$. We can decompose the $4 \times 4$ unitary matrices
$U^{L,R}$ in the block form
\begin{equation}
  U^{L(R)}= \left( \begin{array}{cc}   
   U^{L(R)}_{ee} & U^{L(R)}_{eF}\\
    U^{L(R)}_{Fe} & U^{L(R)}_{FF} \end{array}\right) \, ,
\end{equation}
where $U^{L,R}_{ee}$, $U^{L,R}_{eF}$ and $U^{L,R}_{Fe}$ are $3 \times
3$, $3 \times 1$ and $1 \times 3$ matrices, respectively, and
$U^{L,R}_{FF}$ are complex numbers. We impose that the following
unitary transformation brings $\mathcal{M}_C$ into a block-diagonal
matrix, namely that
\begin{equation} \label{eq:Block}
U^{R^{\dagger}}\left( \begin{array}{cc}   
    m_e & m_\rho \\
   m_S & M_F \end{array} \right) U^{L}= \left( \begin{array}{cc}   
    m_{\rm light} & 0 \\
   0 & m_{\rm heavy} \end{array} \right) \, .
\end{equation}
Eq.~\eqref{eq:Block} imposes constraints on the $U^{L,R}$ matrices and
hence reduces their numbers of independent parameters. In particular,
it requires $U^{L,R}$ to lead to two vanishing $3 \times 1$ and $1
\times 3$ submatrices. Therefore, each of them must have three degrees
of freedom only. We then formulate the ans{\"a}tze for $U^L$ and $U^R$
\begin{align}
U^L &= \left( \begin{array}{cc}   
   \sqrt{\id_3 - LL^{\dagger}}& L\\
    -L^{\dagger} & \sqrt{1-L^{\dagger}L} \end{array}\right) \, , \\
U^R &= \left( \begin{array}{cc}   
   \sqrt{\id_3 - RR^{\dagger}}& R\\
    -R^{\dagger} & \sqrt{1-R^{\dagger}R} \end{array}\right) \, ,
\end{align}
where $\id_3$ is the $3 \times 3$ identity matrix and $L$ and $R$ are
$3 \times 1$ matrices. The $L$ and $R$ matrices must be determined
perturbatively as a function of the parameters in $\mathcal{M}_C$ by
expanding in powers of $1/M_F$. One must also Taylor-expand the square
root. In case of $L$, the expansion is given by
\begin{equation}
  L=L_1+L_2+L_3+ \cdots
\end{equation}
and
\begin{equation}
  \sqrt{\id_3- LL^{\dagger}}=\id_3-\frac{1}{2}LL^{\dagger}-\frac{1}{8}LL^{\dagger}LL^{\dagger}+\cdots
\end{equation}
where the $L_i$ matrices are proportional to $M_F^{-i}$. Analogous
expansions can be given for $R$. The coefficients of the expansion
are computed recursively, imposing that the off-diagonal sub-blocks of
$\mathcal{M}_C$ vanish at each order in $M_F$. Using the
aforementioned hierarchy among scales, this procedure leads to
\begin{align}
U^L &= \left( \begin{array}{cc}   
   \id_3-\frac{1}{2}\frac{m_S^{\dagger} m_S}{M_F^{2}}-\frac{1}{2}\frac{m_e^{\dagger} m_ \rho m_S+ m_S^{\dagger}m_\rho^{\dagger}m_e}{M_F^{3}} & \frac{m_S^{\dagger}}{M_F}+\frac{m_e^{\dagger} m_\rho}{M_F^{2}}-\frac{m_S^{\dagger} m_\rho^{\dagger}m_\rho}{M_F^{3}}\\
 -\frac{m_S}{M_F}-\frac{m_\rho^{\dagger}m_e}{M_F^{2}}+\frac{m_\rho^{\dagger} m_\rho m_S}{M_F^{3}}   & 1-\frac{1}{2}\frac{ m_S m_S^{\dagger}}{M_F^{2}}-\frac{1}{2}\frac{m_S m_e^{\dagger} m_ \rho+ m_\rho^{\dagger}m_e m_S^{\dagger} }{M_F^{3}} \end{array}\right) +\mathcal{O}(M_F^{-3}) \, \label{eq.left}, \\
U^R &= \left( \begin{array}{cc}   
   \id_3-\frac{1}{2}\frac{m_\rho m_\rho^{\dagger}}{M_F^{2}}-\frac{1}{2}\frac{m_\rho m_S m_e^{\dagger}+m_e m_S^{\dagger}  m_\rho^{\dagger}}{M_F^{3}} & \frac{m_\rho}{M_F}+\frac{m_e m_S^{\dagger}}{M_F^{2}}-\frac{m_\rho m_\rho^{\dagger}m_\rho}{2M_F^{3}}\\
   -\frac{m_\rho^{\dagger}}{M_F}-\frac{m_S m_e^{\dagger}}{M_F^{2}}+\frac{m_\rho^{\dagger} m_\rho m_\rho^{\dagger}}{2M_F^{3}}   &  1-\frac{1}{2}\frac{ m_\rho^{\dagger}m_\rho}{M_F^{2}}-\frac{1}{2}\frac{m_S m_e^{\dagger} m_\rho+m_\rho^{\dagger} m_em_S^{\dagger}}{M_F^{3}} \end{array}\right) + \mathcal{O}(M_F^{-3}) \, \label{eq.right},
\end{align}
where we kept only the leading order contribution in $m_\rho$ to the
$\mathcal{O}(M_F^{-3})$ coefficients $L_3$ and $R_3$. The
block-diagonal masses for the light and heavy charged leptons are
finally given by
\begin{align}
  m_{\rm light} &= m_e - \frac{m_\rho m_S}{M_F}-\frac{m_\rho m_\rho^\dagger m_e}{2 \, M_F^2}-\frac{m_e m_S^\dagger m_S}{2 \, M_F^2}+\frac{m_\rho  m_\rho^\dagger m_\rho m_S}{2 \, M_F^3}+\mathcal{O}(M_F^{-3}) \, \label{eq.light}, \\
 m_{\rm heavy} &= M_F + \frac{ m_\rho^\dagger m_\rho}{2 \, M_F}+\frac{m_S m_S^\dagger}{2 \, M_F}+\frac{m_\rho^\dagger m_e m_S^\dagger + m_S m_e^\dagger m_\rho}{2 \, M_F^{2}}-\frac{m_\rho^\dagger m_\rho m_\rho^\dagger m_\rho}{8 \, M_F^3}+\mathcal{O}(M_F^{-3}) \, .
\end{align}
In summary, $m_{\rm light} \approx m_e$ and $m_{\rm heavy} \approx
M_F$, with corrections to these zeroth order results entering at
different orders in $1/M_F$.

%% file: tex/couplings.tex
\section{Majoron couplings}
\label{sec:coups}

In the gauge basis, the interaction terms of the majoron with
neutrinos and charged leptons read
\begin{align}
 -\mathcal{L}_{JNN} &= -\frac{i \, J}{2 \, v \, V^2} \left( \begin{array}{cc}
  \bar{\nu}_L^{c} & \bar{\nu}_R
   \end{array} \right)  \left( \begin{array}{cc}   
    0 & -v_S^2 \, m_D^T \\
    -v_S^2 \, m_D & v^{2} \, M_R \end{array} \right) \, \left( \begin{array}{c}
    \nu_L \\
    \nu_R^{c} \end{array} \right) \, + \hc \, , \label{eq:MajNeutral} \\
 -\mathcal{L}_{JCC} &=-\frac{i \, J}{v \, V^2}  \left( \begin{array}{cc}
  \bar{e}_R & \bar{F}_R
   \end{array} \right)  \left( \begin{array}{cc}   
    v_S^2 \, m_e & v^2 \, m_\rho \\
    -v_H^2 \, m_S & 0 \end{array} \right) \, \left( \begin{array}{c}
    e_L \\
    F_L \end{array} \right) + \hc \, . \label{eq:MajCharged}
\end{align}
The majoron profile in Eq.~\eqref{eq:J} has been used in the
derivation of Eqs.~\eqref{eq:MajNeutral} and \eqref{eq:MajCharged}.
We can now focus on the interaction Lagrangian involving charged
leptons and write it in the fermion mass basis. This results in
\begin{equation}
  -\mathcal{L}_{JCC} = - \frac{i \, J}{v \, V^2} \Big[ \bar{X}^\beta \left(V^{R^{\dagger}}AV^{L}\right)_{\beta\alpha} P_L \, X^\alpha - \bar{X}^\alpha \left(V^{L^{\dagger}}A^{\dagger}V^{R}\right)_{\alpha\beta} P_R \, X^\beta \Big] \, ,
\end{equation}
where $\alpha,\beta$ are flavor indices, specified here for the sake
of clarity, we have defined the four component array in flavor space
$X=(e,F)$ and
\begin{equation} \label{eq:Nmatrix}
  A = \left( \begin{array}{cc}   
    v_S^2 \, m_e & v^2 \, m_\rho \\
    -v_H^2 \, m_S & 0 \end{array} \right)
\end{equation}
is the matrix in Eq.~\eqref{eq:MajCharged}. By comparing to
Eq.~\eqref{eq:llJ}, one finds a dictionary between the $S_{L,R}$
effective couplings and the parameters of the model under
consideration. In the case of the flavor violating couplings, with
$\beta \neq \alpha$, one finds
\begin{align}
  S_L^{\beta\alpha} &= \frac{i}{v \, V^2} \left(V^{R^{\dagger}}AV^{L}\right)^{\beta\alpha} \, , \label{eq:dicSL}  \\
  S_R^{\beta\alpha} &= -\frac{i}{v \, V^2} \left(V^{L^{\dagger}}A^{\dagger}V^{R}\right)^{\beta\alpha} \, . \label{eq:dicSR}
\end{align}
This matching only holds for the light charged leptons, hence $\alpha,\beta = 1, 2, 3$ here. We note that the matching is completely specified by the charged lepton mass matrix $\mathcal{M}_C$. In the case of the flavor conserving couplings, with $\beta=\alpha$, one gets
 \begin{equation} \label{eq:diag}
 S^{\beta\beta} \equiv S_L^{\beta\beta}+S_R^{\beta\beta*}=\frac{i}{v \, V^2} \left(V^{R^{\dagger}}AV^{L}\right)^{\beta\beta} \, .
 \end{equation}
We note that there is a mismatch of a factor of $2$ with the matching
holding for the off-diagonal couplings, which prevents us from writing
a single matching relation. The coupling in Eq.~\eqref{eq:diag} is
purely imaginary, as expected for a pure pseudoscalar. The analytic
proof of this result is given in Appendix~\ref{sec:app1}.

Finally, one can find approximate expressions for the $S_{L,R}$
couplings by using the expressions derived for the $V^{L,R}$ matrices
in the previous Section. One finds
\begin{align}
  S_L^{\beta\alpha} &= \frac{i}{v \, V^2} \, C^{\beta\alpha} \, , \label{eq:SLapp} \\  
  S_R^{\beta\alpha} &= -\frac{i}{v \, V^2} \, {C^{\alpha\beta}}^* \, , \label{eq:SRapp}
\end{align}
with
\begin{equation} \label{eq:C}
  C = D_e^{R^{\dagger}} \, \Big[v_S^2 \, m_e - \frac{v_S^2}{M_F} m_\rho m_S -\frac{3 \, v_S^2+2 \, v_H^2}{2 \, M_F^2} \, m_\rho m_\rho^\dagger m_e+\frac{2 \, v_H^2-v_S^2}{2 \, M_F^2} \, m_e m_S^\dagger m_S + \mathcal{O}(M_F^{-3}) \Big] \, D_e^{L} \, .
\end{equation}
Comparing with Eq.~\eqref{eq.light}, it follows that the off-diagonal
couplings of the Majoron are suppressed by $\mathcal{O}(M_F^{-2})$,
and not by $\mathcal{O}(M_F^{-1})$ as one could naively expect. This
follows from the fact that $D_e^{L,R}$ are the unitary matrices which
diagonalize $m_{\rm light}$.

%% file: tex/pheno.tex
\section{Phenomenology of the model}
\label{sec:pheno}

The model can be probed thanks to its signatures in low-energy flavor
experiments and high-energy colliders.

\subsection*{Lepton flavor violating signatures}

The first and most evident consequence of a massless majoron with
tree-level LFV couplings is the existence of large LFV rates in wide
regions of the parameter space. This includes the usual LFV processes,
such as $\ell_\alpha \to \ell_\beta \, \gamma$ or $\ell_\alpha \to 3
\, \ell_\beta$, which constitute important constraints for our
model. For the radiative processes $\ell_\alpha \to \ell_\beta \,
\gamma$ we use the general formulas in~\cite{Lavoura:2003xp}, whereas
for the 3-body LFV decays $\ell^{-}_{\alpha}\rightarrow
\ell^{-}_{\beta}\ell^{-}_{\beta}\ell^{+}_{\beta}$,
$\ell^{-}_{\alpha}\rightarrow
\ell^{-}_{\beta}\ell^{-}_{\gamma}\ell^{+}_{\gamma}$ and
$\ell^{-}_{\alpha}\rightarrow
\ell^{+}_{\beta}\ell^{-}_{\gamma}\ell^{-}_{\gamma}$, we use the
general expressions in~\cite{Abada:2014kba}. The effective
coefficients for the 3-body decays are generated in our model at
tree-level and are listed in Appendix~\ref{sec:app2}. In addition, we
must consider processes involving the majoron in the final state. In
particular, the LFV decay $\ell_\alpha \to \ell_\beta \, J$ is induced
by the off-diagonal $S_A^{\beta \alpha}$ scalar couplings, with $A =
L,R$, defined in Eq.~\eqref{eq:llJ}. At leading order in $m_\beta /
m_\alpha$, the $\ell_\alpha \to \ell_\beta \, J$ decay width is given
by~\cite{Escribano:2020wua}
\begin{equation}
  \Gamma \left(\ell_\alpha \to \ell_\beta \, J \right) = \frac{m_\alpha}{32 \, \pi} \left( \left| S_L^{\beta \alpha} \right|^2 + \left| S_R^{\beta \alpha} \right|^2 \right) \, .
  \label{eq:decaywidth_betaphi}
\end{equation}

Searches for $\ell_\alpha \to \ell_\beta \, J$ have been performed by
several experiments. The non-observation of these processes has been
used to set stringent limits on the corresponding charged lepton LFV
branching ratios. In the following we will focus on the muon decay $\mu
\to e \, J$. In this case, the strongest limit was obtained at TRIUMF,
finding $\text{BR} \left(\mu \to e \, J\right) < 2.6 \times 10^{-6}$
at 90\% C.L.~\cite{Jodidio:1986mz}. However, the high polarization of
the muon beam used in this experiment implies that the limit is only
strictly valid for a purely right-handed $\mu-e-J$ interaction, with
$S_L^{e \mu} = 0$, as discussed in~\cite{Hirsch:2009ee}. This
reference estimates a more general limit of the order of $\text{BR}
\left( \mu \to e \, J \right) \lesssim 10^{-5}$. A very similar
bound was recently obtained by the TWIST
collaboration~\cite{TWIST:2014ymv}.

\subsection*{Anomalous magnetic moment of the muon}

The enlarged lepton sector in our model induces contributions to many
leptonic observables. We have already discussed flavor violating
observables, which vanish in the SM. In addition, flavor conserving
observables also receive new contributions, and these may potentially
induce deviations from the SM predictions. For instance, the new
states contribute to the anomalous magnetic moment of the muon, an
observable that has received a lot of attention recently.

The anomalous magnetic moments of charged leptons,
\begin{equation}
  a_\alpha = \frac{\left(g-2\right)_\alpha}{2} \, ,
\end{equation}
with $\alpha = e, \mu, \tau$, are described by the effective
Hamiltonian~\cite{Crivellin:2018qmi}
\begin{equation} \label{eq:eff}
\mathcal{H} = c_{\beta\alpha} \, \overline{\ell}_\beta \, \sigma_{\mu \nu} \, P_R \, \ell_\alpha \, F^{\mu \nu} \, , 
\end{equation}
where $F^{\mu \nu}$ is the electromagnetic field strength tensor. One
can obtain the anomalous magnetic moment of the charged lepton
$\ell_\alpha$ as
\begin{equation} \label{eq:relac}
  a_\alpha = - \frac{2 \, m_\alpha}{e} \left( c_{\alpha\alpha} + c_{\alpha\alpha}^\ast \right) = - \frac{4 \, m_\alpha}{e} \, \text{Re} \, c_{\alpha\alpha} \, .
\end{equation}
A discrepancy between the SM prediction for the muon $g-2$ and its
experimentally determined value has existed for a long time. The
interest in this deviation has increased notably after the Muon $g-2$
experiment announced its first results~\cite{Muong-2:2021ojo}. The
combination of their measurement with that obtained by the E821
experiment at Brookhaven~\cite{Muong-2:2006rrc} leads to a $4.2
\sigma$ discrepancy with the SM prediction~\cite{Aoyama:2020ynm},
which can be quantified as
\begin{align} 
  \Delta a_{\mu} = a_\mu^{\text{exp}} - a_\mu^{\text{SM}} = (25.1 \pm 5.9) \times 10^{-10} \, . \label{eq:DeltaaMu}
\end{align}
One should note, however, that a recent calculation of the hadronic
vacuum polarization contribution does not favor such a large
deviation~\cite{Borsanyi:2020mff}. We therefore need additional
experimental data, soon to be provided by the Muon $g-2$
collaboration, as well as more theoretical cross-checks, to firmly
establish the presence of new physics in the muon $g-2$.

\begin{figure}[tb!]
  \centering
  \includegraphics[width=0.6\linewidth]{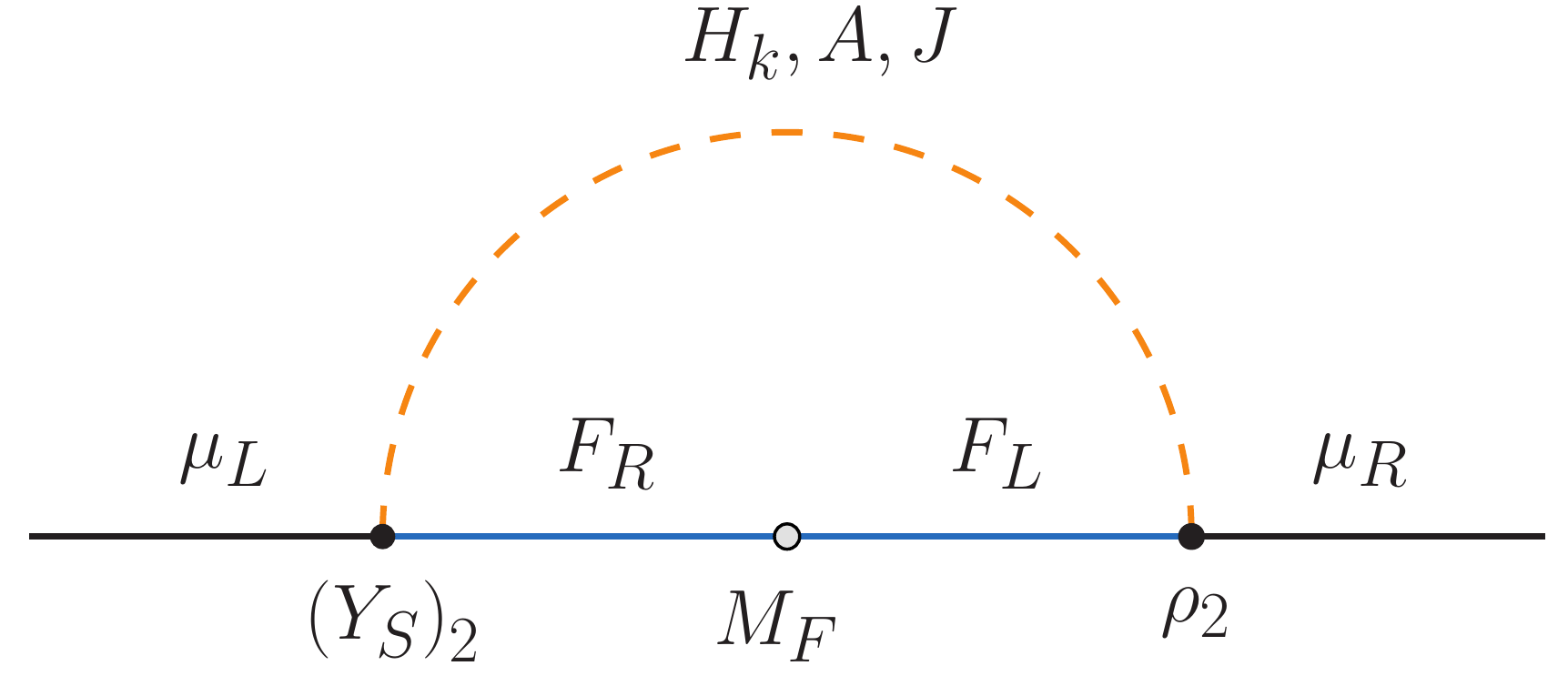}
  \caption{Dominant new physics contribution to the muon anomalous magnetic moment.
    \label{fig:g-2}
    }
\end{figure}

In our model, new contributions to the muon $g-2$ are induced at the
1-loop level, as shown in Fig.~\ref{fig:g-2}. This figure includes
diagrams with the massive CP-even bosons $H_k$, with the massive
CP-odd state $A$ and with the majoron $J$. For the massive states we
use the analytical expressions given in~\cite{Crivellin:2018qmi},
whereas the majoron contribution was recently computed
in~\cite{Escribano:2020wua}.

\subsection*{Higgs boson decays}

The lightest CP-even scalar mass eigenstate, $H_1 \equiv h$, can be
identified with the $125$ GeV state discovered at the LHC. Therefore,
it is crucial that its properties and decay channels match those
observed, within the ranges allowed by the experimental errors. Since
the observed state resembles the Higgs boson of the SM, this is
guaranteed if the mixing angles in the CP-even scalar sector are
sufficiently small. In this case, $h$ is made up mostly by the $S_H$
state, which has the properties of a SM Higgs. Thus, decays such as
$h \rightarrow Z Z$, $W W$ are not modified substantially.

Higgs decays into a pair a of muons have been recently
searched for by the ATLAS~\cite{ATLAS:2020fzp} and
CMS~\cite{CMS:2020xwi} collaborations. Their data yields the following
ratio in terms of the SM predicted
value~\cite{ParticleDataGroup:2020ssz},
\begin{equation} \label{eq:Rhmumu}
R_{h\mu\mu} = \frac{\text{BR}(h \to \mu\mu)^{\rm exp}}{\text{BR}(h \to \mu\mu)^{\rm SM}} = 1.19 \pm 0.39 \, .
\end{equation}
In our model, the mixing in the CP-even scalar sector may induce large
deviations from the SM predicted ratio, $R_{h\mu\mu} =
1$. Even if the mixing is tiny, the large $S$ and $\sigma$ couplings
to muons, required if one wants to address the experimental anomaly in
the muon anomalous magnetic moment, induce sizable contributions to
$R_{h\mu\mu}$, which can largely deviate from $1$. An approximate
analytical expression for $R_{h\mu\mu}$, valid under some
simplifying assumptions, is provided in Appendix~\ref{sec:app3}.

Finally, $h$ can also decay invisibly to a pair of majorons. The
current ATLAS limit on the invisible Higgs branching ratio translates
into $\text{BR}(h \to J J) < 0.11$ at $95\%$
C.L.~\cite{ATLAS:2020kdi}. We take this constraints into account in
our numerical analysis, using $\Gamma_h \approx \Gamma_h^{\rm SM} =
4.1$ MeV for the total Higgs decay
width~\cite{LHCHiggsCrossSectionWorkingGroup:2016ypw}.

%% file: tex/results.tex
\section{Numerical results}
\label{sec:results}

In this Section we present our numerical results. These have been
obtained by randomly scanning in the wide parameter space of the model
and computing the observables discussed in the previous Section. All
points in our scans are compatible with current neutrino oscillation
data. This is achieved by using a Casas-Ibarra
parametrization~\cite{Casas:2001sr} for the $Y_\nu$ Yukawa matrix,
which can be expressed as
\begin{equation} \label{eq:CI}
  Y_\nu = i \, \frac{\sqrt{2}}{v_H} \, D_{\sqrt{M_R}} \, O \, D_{\sqrt{m}} \, U^\dagger \, .
\end{equation}
Here, we are working in the basis in which the $M_R$ matrix, defined
in Eq.~\eqref{eq:Neutrino}, is diagonal, with $M_R =
\text{diag}\left(M_1,M_2,M_3\right)$. The neutrino mass matrix is
diagonalized as $U^T \, m_\nu \, U =
\text{diag}\left(m_1,m_2,m_3\right)$, where $U$ is a $3 \times 3$
unitary\footnote{In a seesaw setup, there will be tiny departures from
unitarity in $U$. For the neutrino fit in Eq.~\eqref{eq:CI} this
effect is numerically irrelevant.}  matrix measured in neutrino
oscillation experiments. We have also defined $D_{\sqrt{M_R}} =
\text{diag}\left(\sqrt{M_1},\sqrt{M_2},\sqrt{M_3}\right)$ and
introduced the $3 \times 3$ orthogonal matrix $O$, such that $O^T O =
O O^T = \id_3$. In our numerical analysis we assume normal neutrino
mass ordering and randomly take neutrino oscillation parameters within
the $3 \, \sigma$ ranges obtained by the global
fit~\cite{deSalas:2020pgw}.

Several parameters are chosen randomly in our scans. These are $v_S$,
$v_\sigma$, the vector-like lepton mass $M_F$, the trilinear coupling
$\mu$ as well as the $\rho$ and $Y_S$ Yukawa couplings, where the
$\rho_2$ and the $\left(Y_S\right)_2$ upper bounds are chosen below
the non-perturbative regime.  In addition, the $\kappa$ Yukawa matrix
has been taken diagonal, with $\kappa_{ii}$ also chosen randomly. All
Yukawa couplings have been assumed to be real for simplicity. The
ranges for the parameters that have been chosen randomly in our scans
are shown on Tab.~\ref{tab:inputs}. We have chosen the VEV $v_S$ in
the narrow range $[0.05,0.1]$ GeV.  This choice is motivated by the
fact that the doublet $S$ does not couple to quarks. A sizable $v_S$
VEV would imply a reduction of $v_H$ and, as a consequence of this, an
increase in the Higgs boson couplings to quarks, already constrained
by LHC data.  Moreover, $v_S$ is also indirectly constrained due
  to its impact on the diagonal majoron couplings, see
  Eq.~\eqref{eq:C}, and a small value is generally
  required. Furthermore, a small value for $v_S$ motivates a similarly
  small value for the $\mu$ trilinear coupling. Otherwise, the heavy
  CP-even scalars $H_k$ and the pseudoscalar $A$ become very heavy and
  their impact on the phenomenology negligible, see
  Eqs.~\eqref{eq:MR2b} and \eqref{eq:mA}. Finally, the scalar
potential parameters $\lambda_\sigma = \lambda_{H S}^{(1)} = 0.1$ and
$\lambda_{S} = \lambda_{H \sigma} = \lambda_{H S}^{(2)} = \lambda_{S
  \sigma} = 0.01$ have been fixed in all the scans. Note, however,
that the exact choice for these quartic couplings is irrelevant for
the observables we study. Also, the small values of $v_S$ and
  the quartics assure that SM precision observables are not substantially
  changed in our model.

%%%%%%%%%%%Inputs%%%%%%%%%%%
{
\renewcommand{\arraystretch}{1.6}
\begin{table}[tb!]
\centering
\begin{tabular}{| c | c |}
\hline
Parameter & Range \\
\hline
$v_S$ & $\left[0.05 \,,\, 0.1\right]$ GeV \\
$v_\sigma$ & $\left[0.75 \,,\, 1.5\right]$ TeV \\
$M_F$ & $\left[0.75 \,,\, 3\right]$ TeV \\
$\mu$ & $\left[0.01 \,,\, 0.1\right]$ GeV \\
$\kappa_{ii}$ & $\left[0.1 \,,\, 1\right]$ \\
$\rho_1$ & $\left[10^{-6} \,,\, 1\right] \times 5 \cdot 10^{-3}$ \\
$\rho_2$ & $\left[5 \cdot 10^{-4} \,,\, \sqrt{4 \pi}\right]$ \\
$\rho_3$ & $\left[0.05 \,,\, 1\right] \times 10^{-7}$ \\
$\left(Y_S\right)_1$ & $\left[0.01 \,,\, 1\right] \times 10^{-6}$ \\
$\left(Y_S\right)_2$ & $\left[0.1 \,,\, \sqrt{4 \pi}\right]$ \\
$\left(Y_S\right)_3$ & $\left[0.001 \,,\, 5\right] \times 10^{-7}$ \\
\hline
\end{tabular}
\caption{Ranges for the parameters that are randomly taken in our
  numerical analysis.
\label{tab:inputs}}
\end{table}
}
%%%%%%%%%%%%%%%%%%%%%%%%%%%%%%%%%

Our analysis has taken into account several experimental bounds. Starting with the LFV processes, we have checked that the branching ratios of the radiative processes $\ell_\alpha \rightarrow \ell_{\beta} \gamma$ as well as the 3-body decays $\ell^{-}_{\alpha}\rightarrow \ell^{-}_{\beta}\ell^{-}_{\beta}\ell^{+}_{\beta}$, $\ell^{-}_{\alpha}\rightarrow \ell^{-}_{\beta}\ell^{-}_{\gamma}\ell^{+}_{\gamma}$ and $\ell^{-}_{\alpha}\rightarrow \ell^{+}_{\beta}\ell^{-}_{\gamma}\ell^{-}_{\gamma}$ satisfy the existing bounds~\cite{ParticleDataGroup:2020ssz}. Regarding the decays $\ell_\alpha \rightarrow \ell_\beta \, J$, we have imposed the restrictions on the flavor violating couplings instead of the branching ratios. To do so, we have defined the combination
\begin{equation}
|S^{\beta\alpha}|=\left( \left| S_L^{\beta\alpha} \right|^2 + \left| S_R^{\beta\alpha} \right|^2 \right)^{1/2} \, 
\end{equation}
and imposed the bounds derived in~\cite{Escribano:2020wua}, that is,
\begin{equation}
|S^{e\mu}| < 5.3 \times 10^{-11} \, , \quad  |S^{e\tau}| < 5.9 \times 10^{-7} \, , \quad  |S^{\mu\tau}| < 7.6 \times 10^{-7} \, .
\end{equation}
On the flavor conserving side, it is well known that astrophysics imposes very stringent constraints on the majoron couplings to charged leptons. These are obtained by considering majoron-induced cooling processes in dense astrophysical media. Regarding the majoron coupling to electrons, Ref.~\cite{Calibbi:2020jvd} finds (at 90\% C.L.)
\begin{equation} \label{eq:See}
  \text{Im} \, S^{e e} < 2.1 \times 10^{-13} \, .
\end{equation}
The majoron coupling to muons has also been studied, but only very recently~\cite{Bollig:2020xdr,Calibbi:2020jvd,Croon:2020lrf}. Using the supernova SN1987A, Ref.~\cite{Croon:2020lrf} finds the limit~\footnote{Ref.~\cite{Croon:2020lrf} also gives the more stringent limit $\text{Im} \, S^{\mu \mu} < 2.1 \times 10^{-10}$, obtained with more \textit{aggressive} assumptions in the simulation of the supernova SN1987A. We have explicitly checked that our conclusions would be the same if one imposes this version of the bound.}
\begin{equation} \label{eq:Smumu}
  \text{Im} \, S^{\mu \mu} < 2.1 \times 10^{-9} \, .
\end{equation}
The impact of this bound on the phenomenology of the model will be studied in detail in the discussion that follows. We have also made sure here that the ratios
\begin{equation}
  R_{Z\ell\ell} = \frac{\Gamma \left( Z \rightarrow \ell^+ \ell^- \right)}{\Gamma_{SM} \left( Z \rightarrow \ell^+ \ell^- \right)} \, ,
\end{equation}
where $\Gamma_{SM} \left( Z \rightarrow \ell^+ \ell^- \right)$ is the SM predicted decay width, lie within the $95\%$ CL range, which is estimated to be $0.995 < R_{Zee} < 1.003$ and $0.993 < R_{Z\mu\mu} < 1.006$~\cite{ParticleDataGroup:2020ssz}. With respect to Higgs decays, we have considered the very recent measurement of the process $h \to \mu \mu$, discussed in the previous Section, and we have rejected the points in our analysis outside the range compatible with Eq.~\eqref{eq:Rhmumu} at $3 \, \sigma$. Finally, points with $\text{BR} \left(h \to JJ \right) > 0.11$~\cite{ATLAS:2020kdi} have been discarded as well. 

\begin{figure}[tb!]
  \centering
  \includegraphics[width=0.6\linewidth]{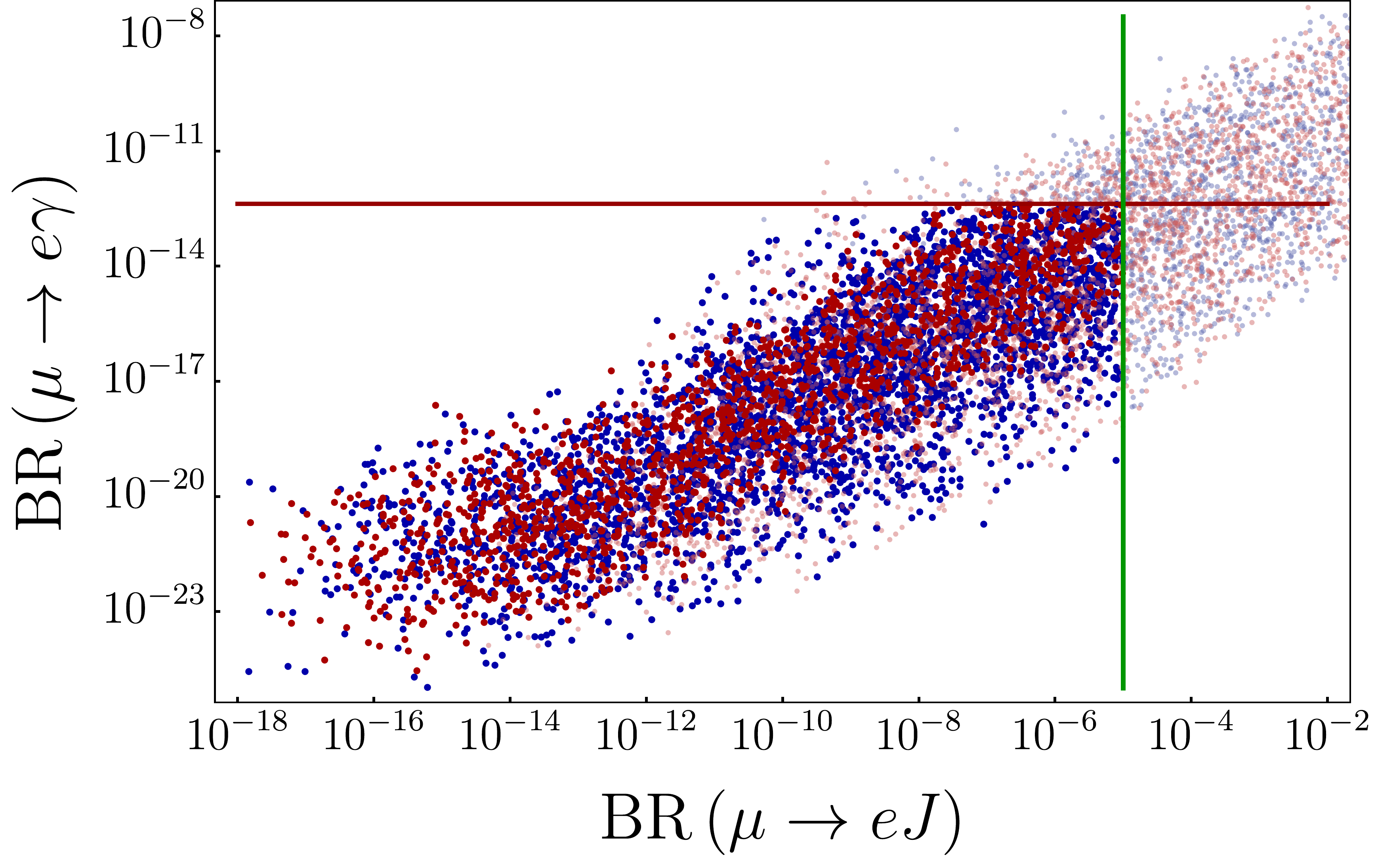}
  \caption{BR($\mu \to e \, J$) as a function of BR($\mu \to e \, J$).
    \label{fig:MuEJ}
    }
\end{figure}

Our results for the LFV processes $\mu \to e \, \gamma$ and $\mu
  \to e \, J$ are shown in Fig.~\ref{fig:MuEJ}, which shows BR($\mu
\to e \, \gamma$) as a function of BR($\mu \to e \, J$). The vertical
line corresponds to the bound BR($\mu \to e \, J$) $< 10^{-5}$,
already discussed in the previous Section, while the horizontal line
is the current limit BR($\mu \to e \, \gamma$) $< 4.2 \times
10^{-13}$, obtained by the MEG experiment~\cite{MEG:2016leq}. Red
  points correspond to parameter points that respect all astrophysical
  bounds, namely the bounds on $S^{ee}$ and $S^{\mu\mu}$ in
  Eqs.~\eqref{eq:See} and \eqref{eq:Smumu}, while the astrophysical
  bound on the majoron coupling to muons is violated in the blue
  points. Finally, the clear points are excluded due to one or several
  of the other experimental constraints mentioned above. First, as can
  be seen in this figure, our model is able to attain the current
  experimental bounds on BR($\mu \to e \, \gamma$) and BR($\mu \to e
  \, J$). Moreover, one finds no difference at all between blue and
  red points. This implies that the astrophysical bounds on the
  flavor-conserving couplings $S^{ee}$ and $S^{\mu\mu}$ have no impact
  on the results for the flavor-violating observables. We also observe
  that a correlation between BR($\mu \to e \, \gamma$) and BR($\mu \to
  e \, J$) exists, although these two observables depend on different
combinations of parameters. However, it is easy to understand that
they are not completely independent. In the limit $\rho_1 = \left( Y_S
\right)_1 = 0$, the vector-like fermion $F$ does not couple to
electrons. In this case, the only contributions to $\mu-e$ LFV
observables come from the $Y_\nu$ Yukawa matrix, which has entries of
the size of $\sim 10^{-7}-10^{-6}$ and then leads to tiny LFV
branching ratios. Therefore, sizable $\rho_1$ or $\left( Y_S
\right)_1$ couplings are required in order to have observable LFV, and
this applies both to $\mu \to e \, J$ and $\mu \to e \,
\gamma$. Regarding other LFV processes, our numerical results also
show that dipole contributions dominate the amplitude of the 3-body
decay $\mu^- \to e^- e^+ e^-$. This leads to strong correlations with
$\mu \to e \, \gamma$, which always has a much larger branching ratio.

Furthermore, in the region of parameter space covered by our
  numerical scan, it is easy to show that BR($\mu \to e \, J$) clearly
  correlates with the combination of parameters $v_\sigma \, \rho_1 \,
  \rho_2 \, M_F^{-2}$. From the expression of the majoron couplings to
  charged leptons in Eqs.~\eqref{eq:SLapp} and \eqref{eq:SRapp} and
  the charged lepton mass matrix in Eq.~\eqref{eq.light}, derived
  under the assumption $m_\rho \ll M_F$, and assuming $v_S \ll v_H$
  and a large $\rho_2$ coupling, as motivated by the explanation of
  the muon $g-2$ anomaly, one can obtain the approximation for the
  off-diagonal $e-\mu$ coupling
\begin{equation} \label{eq:Semuapp}
|S^{e\mu}|\approx\frac{m_\mu \, v_\sigma}{M_F^2} \, \rho_1 \, \rho_2 \, .
\end{equation}
We observe that, as long as the condition $m_\rho < M_F$ is satisfied,
$|S^{e\mu}|$ actually grows when the $\rm U(1)_L$ symmetry breaking
scale $v_\sigma$ increases. This result seems to go against the usual
decoupling behavior expected when the new physics scale becomes
larger. However, when $v_\sigma$ is increased, the mixing between the
SM-like charged leptons and the vector-like lepton $F$ increases as
well, hence enhancing the $\mu-e-J$ coupling. Eventually, when
$v_\sigma$ is pushed above $M_F$, Eq.~\eqref{eq:Semuapp} becomes
invalid and $\text{BR}(\mu \to e \, J)$ starts to decrease.

Several ideas to improve the current limit on $\mu \to e \, J$ have
been put forward recently. As discussed in detail
in~\cite{Perrevoort:2018ttp,Perrevoortthesis}, the limit can be
improved by the Mu3e experiment by looking for a bump in the
continuous Michel spectrum. According to this analysis, $\mu \to e \,
J$ branching ratios above $7.3 \times 10^{-8}$ can be ruled out at
90\% C.L.. Alternatively, reference~\cite{Calibbi:2020jvd} proposes a
new phase of the MEG-II experiment with a Lyso calorimeter in the
forward direction, increasing in this way the sensitivity for $\mu \to
e \, J$. Therefore, $\mu \to e \, J$ already excludes a region of the
parameter space of the model, and this region will be substantially
enlarged in the future.

In what concerns $\tau$ decays, our choice of parameters suppresses
all the LFV amplitudes. Since experimental limits in the $\tau$ sector
are much weaker than for the muon, we do not show plots for LFV $\tau$
decays. We note, however, that one can saturate (some of) the
experimental bounds also for $\tau$'s in our model, for the
appropriate choice of (large) parameters in the 3rd generation. On the
other hand, in our model it is not possible to have both, $\tau\to e
\, \gamma$ and $\tau\to \mu \, \gamma$, with large rates at the same
time, without running into conflict with $\mu \to e \, \gamma$.

\begin{figure}[tb!]
  \centering
  \includegraphics[width=0.6\linewidth]{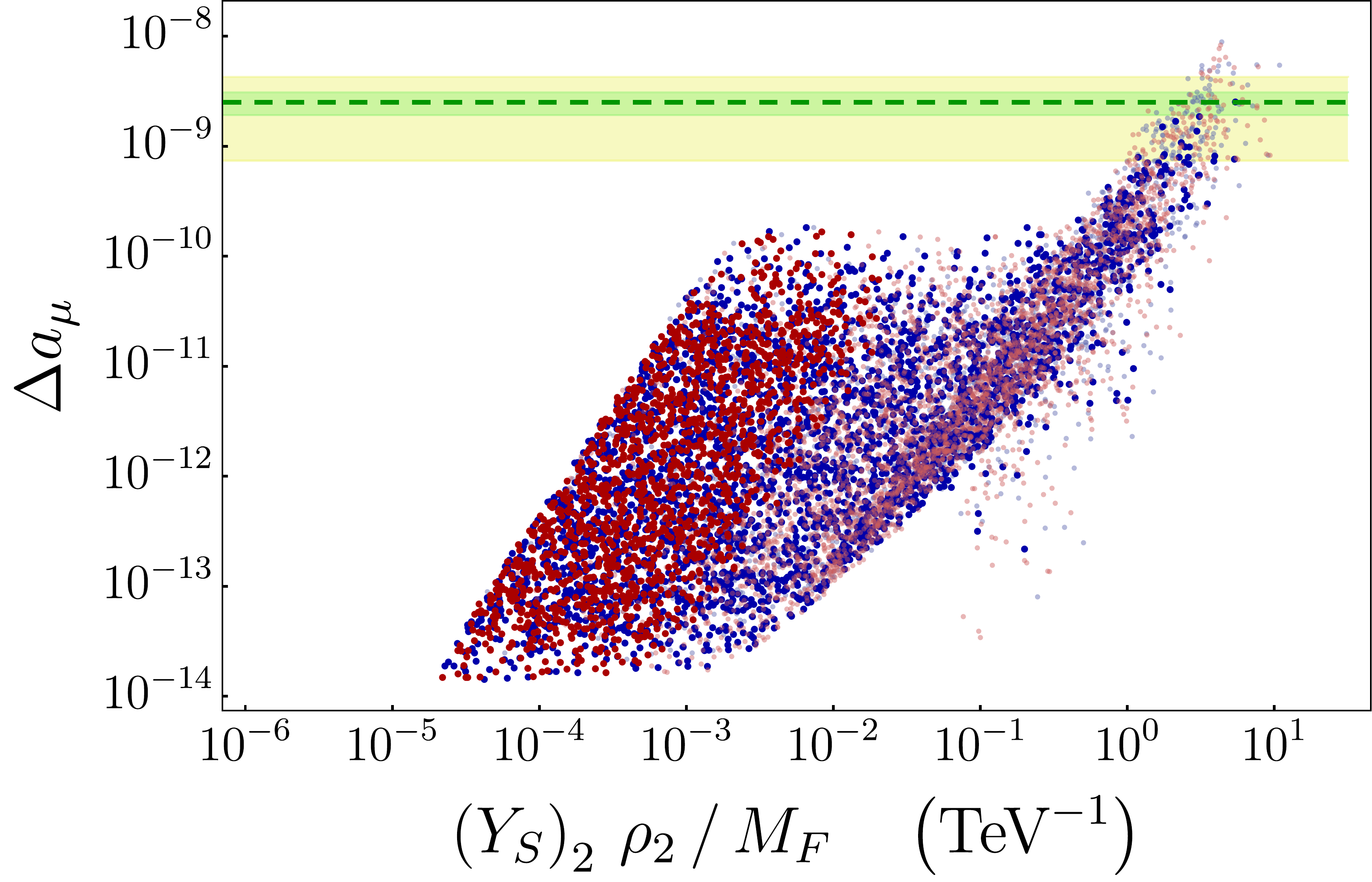}
  \caption{$\Delta a_\mu$ as a function of the combination
    $\left(Y_S\right)_2 \rho_2 / M_F$. Gray points are excluded due to
    one or several experimental bounds, but are shown for
    illustration.
    \label{fig:Delta1}
    }
\end{figure}

\begin{figure}[tb!]
  \centering
  \includegraphics[width=0.6\linewidth]{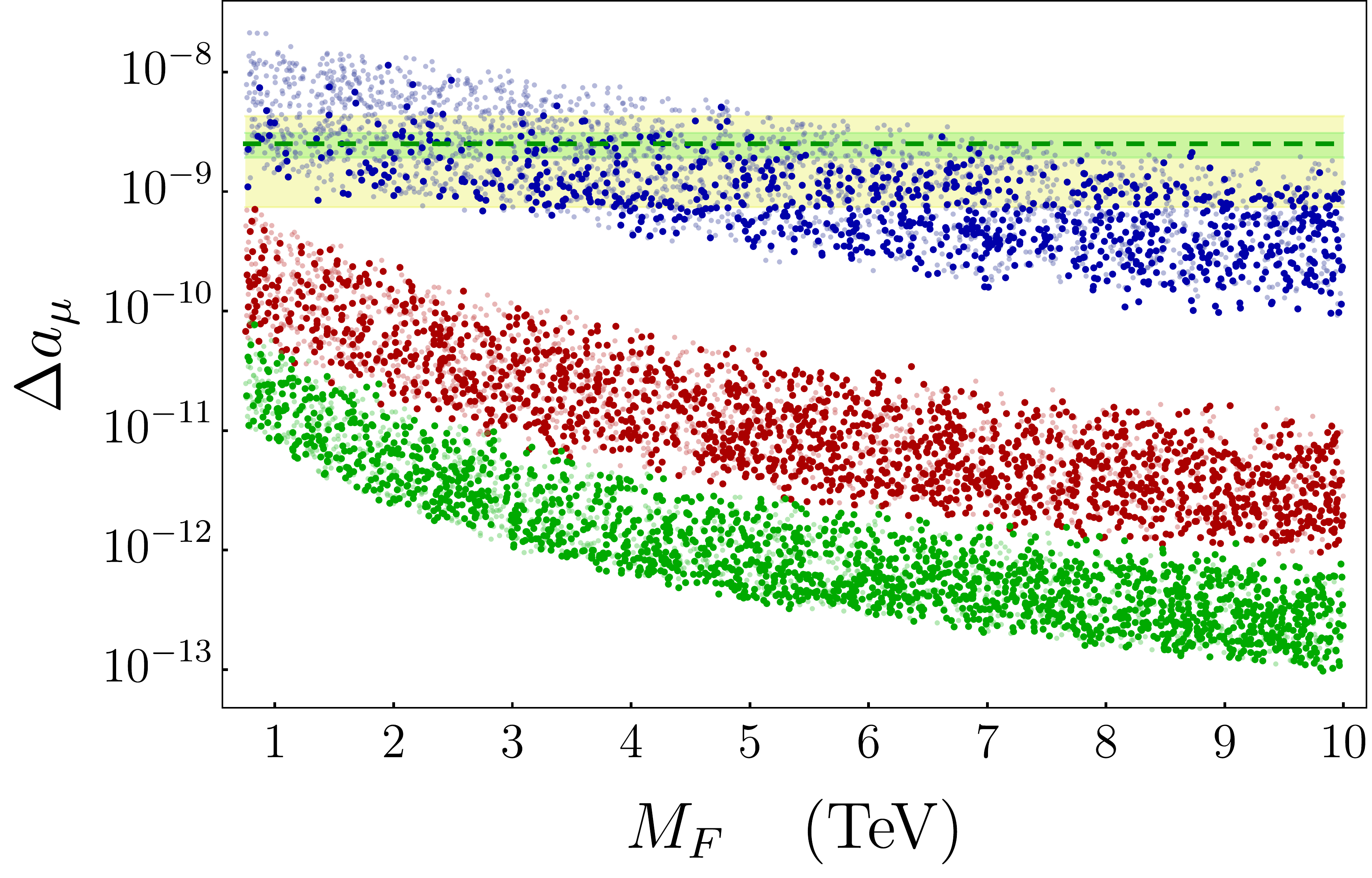}
  \caption{$\Delta a_\mu$ as a function of the vector-like mass $M_F$
    for $\rho_2 = \left(Y_S\right)_2 = \sqrt{4 \pi}$ (blue), $\rho_2 =
    \left(Y_S\right)_2 = 1$ (red) and $\rho_2 = \left(Y_S\right)_2 =
    0.5$ (green).
    \label{fig:Delta2}
    }
\end{figure}

Our model can also induce large contributions to the muon anomalous
magnetic moment and address the current experimental anomaly. This is
shown in Fig.~\ref{fig:Delta1}. This figure displays $\Delta a_\mu$ as
a function of the combination of parameters $\left(Y_S\right)_2 \rho_2
/ M_F$, which enters the Feynman diagram in Fig.~\ref{fig:g-2}. The
horizontal dashed line represents the experimental central value,
while the green and yellow bands correspond to the $1 \sigma$ and $3
\sigma$ ranges, respectively.~\footnote{We note that the $1 \sigma$
and $3 \sigma$ intervals are symmetric with respect to the central
value, but they look asymmetric in this figure since we are using a
logarithmic scale for the y-axis.} As in the previous figure, the
  red points respect the astrophysical bounds on $S^{ee}$ and
  $S^{\mu\mu}$, while the blue points only respect the constraint on
  $S^{ee}$. Clear points are excluded due to one or several
  constraints, but are shown for illustration. We have found
numerically that all diagrams, with massive scalars or with the
majoron in the loop, may have comparable sizes. Interestingly,
some points are found within the $1 \sigma$ interval, hence providing
a good explanation for the experimental value of the muon anomalous
magnetic moment. These points require relatively light $F$ fermions
(with masses of the order of $\sim 1-2$ TeV) and large (order $1$)
$\rho_2$ and $\left(Y_S\right)_2$ Yukawa couplings. However, they
  violate the bound on $S^{\mu\mu}$ obtained from the supernova
  SN1987A, since this constraint necessarily implies a low value of
  $\rho_2$. In fact, we note that this figure displays a large
  concentration on points with low values of $\Delta a_\mu$ in a
  region where all the red points are found. This region is
  characterized by $\rho_2 \ll 1$, and hence the dominant
  contributions to the muon $g-2$ do not come from the diagram in
  Fig.~\ref{fig:g-2}, but are mostly induced by diagrams proportional
  to $\left(Y_S\right)_2^2 / M_F^2$. These diagrams have an external
  chirality flip that introduces an $m_\mu$ suppresing factor and
  then, as is generically found in a large class of models with this
  feature, $\Delta a_\mu$ can be at most $\sim 10^{-10}$.

Complementary information is provided by Fig.~\ref{fig:Delta2}, which
shows $\Delta a_\mu$ as a function of the vector-like mass $M_F$
for three different values of $\rho_2 = \left(Y_S\right)_2$. Blue
  points corresponds to $\rho_2 = \left(Y_S\right)_2 = \sqrt{4 \pi}$,
  red points to $\rho_2 = \left(Y_S\right)_2 = 1$ and green points to
  $\rho_2 = \left(Y_S\right)_2 = 0.5$. This figure has been obtained
with a specific parameter scan in which $M_F \in \left[ 0.75 \,,\,
    10 \right]$ TeV, while the ranges for the other randomly chosen
  parameters are as in Tab.~\ref{tab:inputs}. As expected, all new
physics contributions decrease for large $M_F$ and strongly depend
  on the value of the $\rho_2$ and $\left(Y_S\right)_2$
  couplings. When $\rho_2 = \left(Y_S\right)_2 = 0.5$, these are not
  large enough to address the muon $g-2$ anomaly, while when $\rho_2 =
  \left(Y_S\right)_2 = 1$ this happens in a narrow region of the
  parameter space characterized by very light vector-like leptons,
  with masses $\lesssim 1$ TeV. Only when $\rho_2 = \left(Y_S\right)_2
  = \sqrt{4 \pi}$, one can find an explanation for the anomaly in a
  wide $M_F$ range. And even in this case, they eventually become too
  small to account for the measured muon $g-2$. However, this happens
for very large vector-like masses. In fact, one finds that vector-like
masses as large as $10$ TeV still allow for a $3 \sigma$ explanation
of the muon $g-2$ anomaly. Such a large mass would make the $F$
fermions unobservable at the LHC.

\begin{figure}[tb!]
  \centering
  \includegraphics[width=0.6\linewidth]{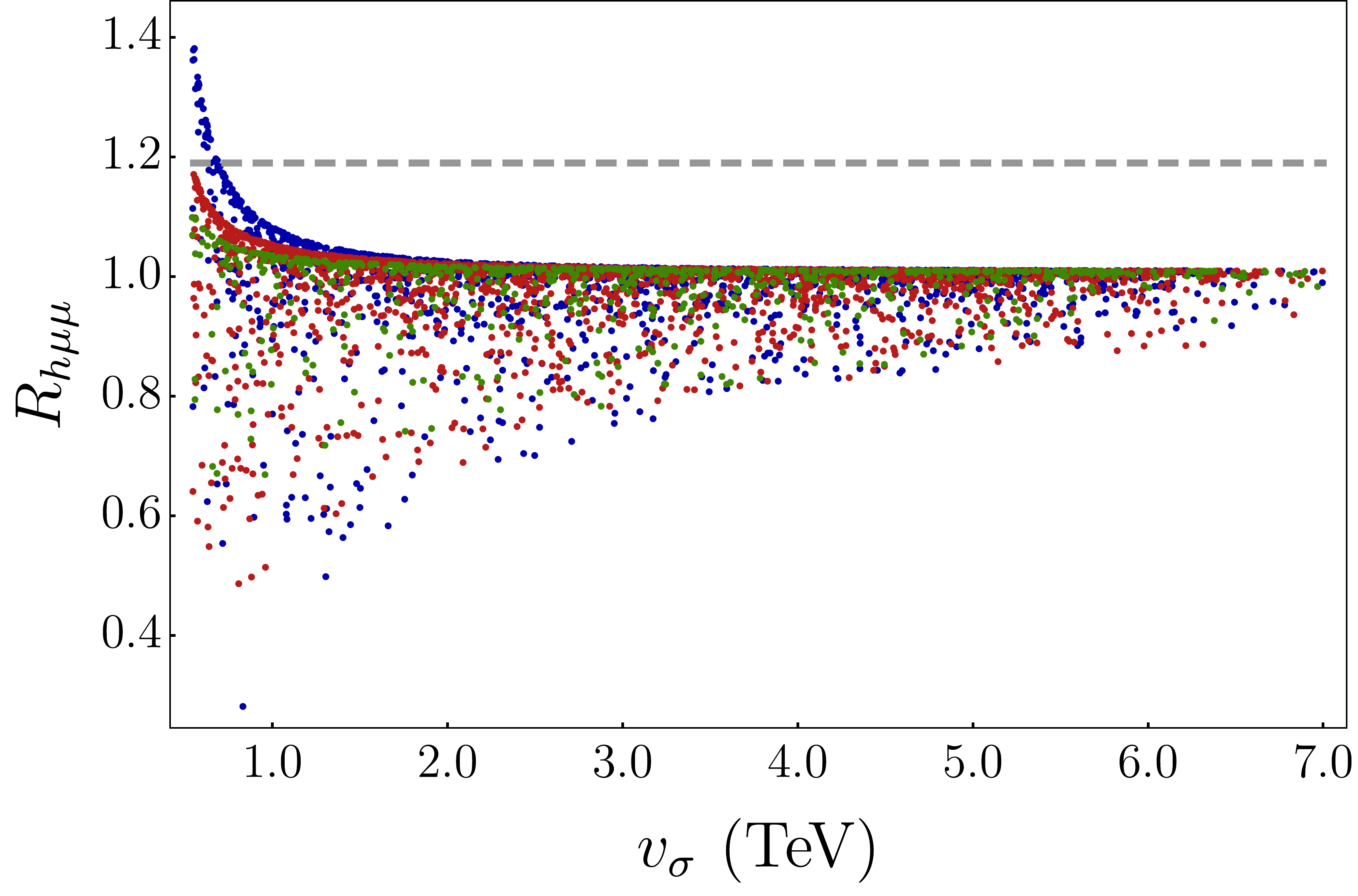}
  \caption{$R_{h\mu\mu}$ as a function of $v_\sigma$ for three fixed
    values of the vector-like mass $M_F$: $M_F = 1$ TeV (blue), $M_F =
    3$ TeV (red) and $M_F = 5$ TeV (green). The horizontal line
    represents the current central value, $R_{h\mu\mu} =
    1.19$~\cite{ParticleDataGroup:2020ssz}.
    \label{fig:Rhmumu}
    }
\end{figure}

We turn our attention to Higgs boson decays. As already explained, the
mixing in the CP-even scalar sector can induce large deviations from
the SM predicted Higgs branching ratios. In particular, a large
effective coupling to muons is induced in parameter points in which
the muon $g-2$ anomaly is explained. This is shown in
Fig.~\ref{fig:Rhmumu}. Here we plot the ratio $R_{h\mu\mu}$, defined
in Eq.~\eqref{eq:Rhmumu}, as a function of the lepton number breaking
scale $v_\sigma$. The horizontal line represents the current central
value, $R_{h\mu\mu} = 1.19$~\cite{ParticleDataGroup:2020ssz}. As
  in the previous plot, the vector-like mass $M_F$ is fixed to
specific values in this figure: $M_F = 1$ TeV (blue points), $M_F = 3$
TeV (red points) and $M_F = 5$ TeV (green points). In addition,
$v_S = 0.1$ GeV and $\rho_2 = \left(Y_S\right)_2 = \sqrt{4\pi}$
  are fixed in this plot, while $\mu \in \left[0.05 \,,\, 50\right]$
  GeV, $\rho_1 \in \left[0.002 \,,\, 1.2\right] \times 10^{-6}$ and
  $\left(Y_S\right)_1 \in \left[0.1 \,,\, 5\right] \times 10^{-7}$ are
  randomly varied and the rest of parameters are taken as in
  Tab.~\ref{tab:inputs}. Due to the large value chosen for $\rho_2$,
  the astrophysical bound on $S^{\mu\mu}$ is not respected in this
  plot. Imposing this constraint would imply $R_{h\mu\mu} \approx 1$.
We observe that for large $v_\sigma$ and $M_F$ the new physics
contributions become negligibly small and one finds $R_{h\mu\mu} =
1$. However, for lower scales one finds many parameter points leading
to large deviations from the SM predicted value. In particular, for
$M_F = 1$ TeV our scan reveals points with $R_{h\mu\mu}$ as large
$\sim 1.4$ or as low as $\sim 0.3$. These extreme points are of course
ruled out by the existing data, but serve as example of how easily
Higgs decays into muons can deviate from the SM predictions in our
setup.

We finally note that the $R_{h\mu\mu}$ ratio does not correlate with
other observables, due to the large number of independent
contributions to the $h-\mu-\mu$ coupling, see
Appendix~\ref{sec:app3}. For this reason, a definite prediction cannot
be made. For instance, Eqs.~\eqref{eq:cmumu1}-\eqref{eq:cmumu3} imply
that for vanishing mixing in the scalar sector,
$c_{\mu\mu}^{S_{\sigma}} = c_{\mu\mu}^{S_S} = 0$ and $c_{\mu\mu}^{S_H}
> c_{\mu\mu}^{\rm SM}$, hence predicting $R_{h\mu\mu} > 1$. However,
the $\alpha$ and $\beta$ angles never vanish and in fact one can find
$R_{h\mu\mu} < 1$ as well.

%% file: tex/conclusions.tex
\section{Summary}
\label{sec:conclusions}

In this paper we have proposed a simple model that leads to sizable
majoron flavor violating couplings to charged leptons. The particle
spectrum is extended with the addition of two new scalar multiplets,
as well as three right-handed neutrino singlets and a vector-like
lepton. The SM symmetry is also extended with a continuous lepton
number global symmetry. As a result of spontaneous symmetry breaking,
neutrinos acquire non-zero Majorana masses via a type-I seesaw
mechanism and a massless Goldstone boson appears in the spectrum, the
majoron.

Thanks to large mixings between the SM charged leptons and the
vector-like lepton, sizable majoron LFV couplings are generated at
tree-level. Therefore, our model constitutes a simple example of a
model with tree-level off-diagonal majoron couplings, not suppressed
by neutrino masses. This induces plenty of signatures in experiments
looking for LFV processes. In particular, we have shown that the decay
$\mu \to e \, J$ can have large rates, close to the experimental
limit. In fact, it already excludes part of the parameter space of the
model.

As a by-product of our construction, other interesting
phenomenological possibilities emerge: (i) an explanation to the
current muon $g-2$ discrepancy can be provided in large parts of the
parameter space of the model, easily finding points that address the
anomaly even within $1 \sigma$, and (ii) sizable deviations with
respect to the SM predicted Higgs decay rates can be obtained, most
notably in $h \to \mu \mu$. These two phenomenological possibilities
provide additional handles on the model. We note, however, that an
  explanation of the muon $g-2$ anomaly would lead to tension with
  recent astrophysical bounds on the majoron coupling to muons.

In this work we have shown that as soon as the lepton sector is
extended beyond the minimal models, exotic signatures appear, such as
those including a massless majoron in the final state. This motivates
the experimental search for processes like $\mu \to e \, J$ and
estimulates the construction of new theoretical constructions that, in
addition to neutrino masses, provide an understanding to other open
questions.

%% file: tex/app1.tex
\section{Proof of the pseudoscalar nature of the majoron couplings}
\label{sec:app1}

Eq.~\eqref{eq:diag} encodes the diagonal couplings of the majoron with
the charged leptons of our model. Since the majoron is a pure
pseudoscalar Goldstone boson, the coefficients $S^{\beta\beta}$ are
purely imaginary. We are going to prove that this is indeed the case
for the general scenario of $n$ singlet vector-like lepton pairs added
to the SM leptons, with the same form of the couplings as the one
defined in Eq.~\eqref{eq:Yextra}.~\footnote{We thank Isabel
Cordero-Carri\'on for providing the seed for this proof.} Let
\begin{equation} M = \left( \begin{array}{cc}   
    m_1 & m_2 \\
    m_3 & m_4 \end{array} \right) \, 
\end{equation}
be a generic complex $(3+n)\times (3+n)$ complex matrix, given by the
blocks $m_1$, $m_2$, $m_3$ and $m_4$, with dimensions $3 \times 3$, $3
\times n$, $n \times 3$ and $n \times n$, respectively. The singular
value decomposition of the matrix $M$ is $V_R^{\dagger} \, M \, V_L =
\widehat{M}$, where $V_R$ and $V_L$ are unitary matrices and
$\widehat{M}=\text{diag}(M_1, M_2)$, with $M_1$ and $M_2$ $3 \times 3$
and $n\times n$ real diagonal matrices, respectively, with positive
entries. The interaction matrix of the majoron with the charged
leptons can be written in the flavor basis as
\begin{equation}
N = \left( \begin{array}{cc}   
     x\,m_1 & (x+y)\,m_2 \\
    -y\,m_3 & 0 \end{array} \right) \, ,
\end{equation}
where $N$ is another $(3+n)\times (3+n)$ matrix and $x,y \in
\mathbb{R}$. Comparing with Eq.~\eqref{eq:Nmatrix}, the majoron
coupling matrix in our model is given by $N \equiv A$ and corresponds
to $m_1 \equiv m_e$, $m_2 \equiv m_\rho$, $m_3 \equiv m_S$, $x \equiv
v_S^2$ and $y \equiv v_H^2$.

First of all, we block-parametrize the unitary matrices $V_R$ and
$V_L$ as
\begin{equation}
 V_R^{\dagger}= \left( \begin{array}{cc}   
    A & B \\
    C & D \end{array} \right) \, , \quad V_L^{\dagger} = \left( \begin{array}{cc}   
    E & F \\
    G & H \end{array} \right) \, .
\end{equation}
We also denote
\begin{equation} 
 e_1=\left( \begin{array}{c}
\id_3 \\ 0 \end{array}\right) \, , \quad e_2=\left( \begin{array}{c}
0 \\ \id_n \end{array}\right) \, ,
\end{equation}
where $\id_n$ is the $n \times n$ identity matrix. Therefore, one obtains
\begin{equation}\begin{split}
\left(V_R^\dagger \, e_1\right) \left(V_R^\dagger \, e_1\right)^{\dagger}=\left( \begin{array}{cc}   
    AA^{\dagger} & AC^{\dagger} \\
    CA^{\dagger} & CC^{\dagger} \end{array} \right) \, , \\
\left(V_R^\dagger \, e_2\right) \left(V_R^\dagger \, e_2\right)^{\dagger}=\left( \begin{array}{cc}   
    BB^{\dagger} & BD^{\dagger} \\
    DB^{\dagger} & DD^{\dagger} \end{array} \right) \, .
\end{split} \end{equation}
Analogously, one finds the following relations involving $V_L$:
\begin{equation}\begin{split}
\left(V_L^\dagger \, e_1\right) \left(V_L^\dagger \, e_1\right)^{\dagger}=\left( \begin{array}{cc}   
    EE^{\dagger} & EG^{\dagger} \\
    GE^{\dagger} & GG^{\dagger} \end{array} \right)  \,, \\
\left(V_L^{\dagger}e_2\right)\left(V_L^{\dagger}e_2\right)^{\dagger}=\left( \begin{array}{cc}   
    FF^{\dagger} & FH^{\dagger} \\
    HF^{\dagger} & HH^{\dagger} \end{array} \right) \, .
\end{split} \end{equation}

After these preliminaries, we note that the interaction matrix $N$ can
be written as
\begin{equation}
N = x \, \left( \begin{array}{cc}   
    m_1 & m_2 \\
    0 & 0 \end{array} \right) + y \, \left( \begin{array}{cc}   
    0 & m_2 \\
    -m_3 & 0 \end{array} \right) \, . \label{eq:decomposition}
\end{equation}
Therefore, if we prove that the combinations
\begin{equation}
V_R^{\dagger}\left( \begin{array}{cc}   
    m_1 & m_2 \\
    0 & 0 \end{array} \right) V_L  \, , \quad  V_R^{\dagger}\left( \begin{array}{cc}   
    0 & m_2 \\
    -m_3 & 0 \end{array} \right) V_L
\end{equation}
have real diagonal elements, then also $V_R^\dagger N V_L$ has real
diagonal elements and the proof is complete. Let us consider the first
term. It is easy to check that
\begin{equation}
\left( \begin{array}{cc}   
    m_1 & m_2 \\
    0 & 0 \end{array} \right)=e_1e_1^{T}M \, .
\end{equation}
Then one can obtain
\begin{equation} \begin{split}
V_R^{\dagger}\left( \begin{array}{cc}   
    m_1 & m_2 \\
    0 & 0 \end{array} \right) V_L=V_R^{\dagger}e_1e_1^{T}MV_L=V_R^{\dagger}e_1e_1^{T}V_R\widehat{M}=\\
\left(V_R^{\dagger} \, e_1\right)\left(V_R^{\dagger} \, e_1\right)^{\dagger}\widehat{M}=\left( \begin{array}{cc}   
    AA^{\dagger}M_1 & AC^{\dagger}M_2 \\
    CA^{\dagger}M_1 & CC^{\dagger}M_2 \end{array} \right) \, .
\end{split}\end{equation}
The diagonal terms of the resulting matrix are real because
$AA^{\dagger}$ and $CC^{\dagger}$ are Hermitian matrices and $M_1,
M_2$ are real diagonal matrices. We now have to consider the second
term in Eq.~\eqref{eq:decomposition}. It is possible to write
\begin{equation}\left( \begin{array}{cc}   
    0 & m_2 \\
    -m_3 & 0 \end{array} \right) =e_1\left(e_1^{T}Me_2\right)e_2^{T}-e_2\left(e_2^{T}Me_1\right)e_1^{T} \, .
\end{equation}
Using similar manipulations as for the first term one finds
\begin{equation}
 V_R^{\dagger}\left( \begin{array}{cc}   
    0 & m_2 \\
    -m_3 & 0 \end{array} \right) V_L=\left(V_R^{\dagger} \, e_1\right)\left(V_R^{\dagger} \, e_1\right)^{\dagger}\widehat{M}\left(V_L^{\dagger} \, e_2\right)\left(V_L^{\dagger} \, e_2\right)^{\dagger}-\left(V_R^{\dagger} \, e_2\right)\left(V_R^{\dagger} \, e_2\right)^{\dagger}\widehat{M}\left(V_L^{\dagger} \, e_1\right)\left(V_L^{\dagger} \, e_1\right)^{\dagger} \, ,
\end{equation}
and, writing for the sake of brevity only the diagonal blocks of this
expression, in the form of a column array, we obtain
\begin{equation}
 \Big[V_R^{\dagger}\left( \begin{array}{cc}   
    0 & m_2 \\
    -m_3 & 0 \end{array} \right) V_L\Big]_{\rm diag} =
\left( \begin{array}{c}   
    AA^{\dagger}M_1FF^{\dagger}+AC^{\dagger}M_2HF^{\dagger} \\ CC^{\dagger}M_2 HH^{\dagger}+ CA^{\dagger}M_1FH^{\dagger}\end{array} \right)-\left( \begin{array}{c}   
    BB^{\dagger}M_1EE^{\dagger}+BD^{\dagger}M_2GE^{\dagger} \\ DD^{\dagger}M_2GG^{\dagger}+DB^{\dagger}M_1EG^{\dagger} \end{array} \right) \,.
\end{equation}
The second terms in the sum cancel for both the upper and lower diagonal blocks, using the unitarity of $V_L$ and $V_R$. Then, using again unitarity in the following way
\begin{equation}\begin{cases}
AA^{\dagger}=1-BB^{\dagger}\\
EE^{\dagger}=1-FF^{\dagger}\\
CC^{\dagger}=1-DD^{\dagger}\\
GG^{\dagger}=1-HH^{\dagger} \, ,
\end{cases}\end{equation}
we finally end up with
\begin{equation} \Big[V_R^{\dagger}\left( \begin{array}{cc}   
    0 & m_2 \\
    -m_3 & 0 \end{array} \right) V_L\Big]_{\rm diag}=\left( \begin{array}{c} M_1FF^{\dagger}-BB^{\dagger}M_1 \\ M_2HH^{\dagger}-DD^{\dagger}M_2  \end{array} \right) \, ,
\end{equation}
and therefore the diagonal components of this matrix are purely real. This concludes the proof.

%% file: tex/app2.tex
\section{Effective coefficients for flavor violating observables}
\label{sec:app2}

In order to use the analytical results for the flavor violating
observables provided in~\cite{Abada:2014kba,Escribano:2020wua}, one
must match the effective Lagrangian in these references to the
specific model discussed here. We focus in particular on the 3-body
lepton decays $\ell^{-}_{\alpha}\rightarrow
\ell^{-}_{\beta}\ell^{-}_{\beta}\ell^{+}_{\beta}$,
$\ell^{-}_{\alpha}\rightarrow
\ell^{-}_{\beta}\ell^{-}_{\gamma}\ell^{+}_{\gamma}$ and
$\ell^{-}_{\alpha}\rightarrow
\ell^{+}_{\beta}\ell^{-}_{\gamma}\ell^{-}_{\gamma}$. These processes
get tree-level contributions in our model from the three CP-even
scalars $H_k$ ($k=1,2,3$), the $Z$-boson, the CP-odd scalar $A$ and
the majoron $J$. The majoron contributions have been computed
in~\cite{Escribano:2020wua}. Since all the other mediators are
significantly heavier than the SM charged leptons, we can then
parametrize their contributions by the effective Lagrangian
\begin{equation} \label{eq:4lepton}
\mathcal{L}_{4\ell}=\sum_{\substack{I=S,V,T \\ X,Y=L,R}}A^{I}_{XY}\bar{\ell^{\beta}}\Gamma_IP_X \ell^{\alpha}\bar{\ell^{\delta}}\Gamma_I P_Y \ell^{\gamma} + \hc \, , 
\end{equation}
where we have defined $\Gamma_S=1$, $\Gamma_V=\gamma_{\mu}$ and
$\Gamma_T=\sigma_{\mu\nu}$ and omitted flavor indices in the effective
coefficients for the sake of simplicity. The coefficients $A^{I}_{XY}$
have dimensions of mass$^{-2}$. We will now give specific expressions
for these coefficients in the model under consideration. In order to
do that it proves convenient to define the following 3-component array
\begin{equation}
(c_1 ,c_2 ,c_3)^{\beta\alpha}=(V^{R^{\dagger}}_{ee}Y_eV^{L}_{ee},V^{R^{\dagger}}_{ee}\rho V^{L}_{Fe} ,V^{R^{\dagger}}_{Fe}Y_S V^{L}_{ee})^{\beta\alpha} \, ,
\end{equation}
which encodes the interactions of the SM charged leptons with the
CP-even scalars gauge eigenstates $\{S_H, S_{\sigma}, S_S\}$.

\subsection*{$\ell^{-}_{\alpha}\rightarrow \ell^{-}_{\beta}\ell^{-}_{\beta}\ell^{+}_{\beta}$}

Recalling the definition of the $3 \times 3$ unitary matrix $W$ given
in Eq.~\eqref{eq:CPeven}, we get the following expressions for the
effective coefficients
$A^{I}_{XY}=(A^{I}_{XY})^{\beta\beta\beta\alpha}$.

\subsubsection*{$H_k$ contributions}

\begin{equation}
A_{LL}^{S}=\frac{1}{2m_{H_k}^{2}}\sum_{i,j}\big(W_{ki}c_i^{\beta\beta}\big)\big(W_{kj}c_j^{\beta\alpha}\big)
\end{equation}

\begin{equation}
A_{LR}^{S}=\frac{1}{2m_{H_k}^{2}}\sum_{i,j}\big(W_{ki}c_i^{\beta\beta}\big)\big(W_{kj} c_j^{\dagger^{\beta\alpha}}\big)
\end{equation}

\begin{equation}
A_{RL}^{S}=\frac{1}{2m_{H_k}^{2}}\sum_{i,j}\big(W_{ki}c_i^{\dagger^{\beta\beta}}\big)\big(W_{kj}c_j^{\beta\alpha}\big)
\end{equation}

\begin{equation}
A_{RR}^{S}=\frac{1}{2m_{H_k}^{2}}\sum_{i,j}\big(W_{ki}c_i^{\dagger{\beta\beta}}\big)\big(W_{kj}c_j^{\dagger{\beta\alpha}}\big)
\end{equation}

\subsubsection*{$Z$ contributions}

\begin{equation}
A_{LL}^{V}=-\frac{g^{2}}{4 m_W^{2}}\Big(V^{L^{\dagger}}_{ee}V^{L}_{ee}-2\sin^{2}\theta_W\mathbb{I}\Big)^{\beta\beta}\Big(V^{L^{\dagger}}_{ee}V^{L}_{ee}\Big)^{\beta\alpha}
\end{equation}
\begin{equation}
A_{LR}^{V}=0
\end{equation}\begin{equation}
A_{RL}^{V}=\frac{g^{2}}{2 m_W^{2}}\sin ^{2}\theta_W\Big(V^{L^{\dagger}}_{ee}V^{L}_{ee}\Big)^{\beta\alpha}
\end{equation}\begin{equation}
A_{RR}^{V}=0
\end{equation}

Actually, the parametrization $V^{L(R)}=U^{L(R)}D^{L(R)}$ greatly
simplifies the expressions. Taking into account the unitarity of
$D^{L}_e$ one ends up with
\begin{equation}
A_{LL}^{V}=\frac{g^{2}}{4 m_W^{2}M_F^{2}}\Big[1-2\sin^{2}\theta_W-\Big(D^{L^{\dagger}}_e \frac{m_S^{\dagger}m_S}{M_F^{2}}D^{L}_e\Big)^{\beta\beta}\Big]\Big(D^{L^{\dagger}}_e m_S^{\dagger}m_S D^{L}_e\Big)^{\beta\alpha}
\end{equation}

\begin{equation}
A_{RL}^{V}=-\frac{g^{2}}{2 m_W^{2}M_F^{2}}\sin^{2}\theta_W \Big(D^{L^{\dagger}}_em_S^{\dagger}m_S D^{L}_e\Big)^{\beta\alpha}
\end{equation}

\subsubsection*{$A$ contributions}

From the profile of the massive CP-odd state $A$ given in Eq.~\eqref{eq:A} one can recover the interaction Lagrangian between $A$ and the charged leptons in the flavor basis
\begin{equation}
 \mathcal{L}_{A\ell\ell}=-\frac{i \, A}{\sqrt{2} V^2}  \left( \begin{array}{cc}
  \bar{e}_R & \bar{F}_R
   \end{array} \right)  \left( \begin{array}{cc}   
    -Y_e \, v_S v_{\sigma}  & \rho \,  v_H v_S \\
    Y_S \, v_H v_{\sigma} & 0 \end{array} \right) \, \left( \begin{array}{c}
    e_L \\
    F_L \end{array} \right) + \hc \, .
\end{equation}
Denoting the matrix in the previous equation as $B$ and transforming
the Lagrangian to the mass basis, one can easily perform the matching
with Eq.~\eqref{eq:4lepton}. We get the following expressions for the
contributions of $A$ to the effective coefficients
$A_{XY}^{I}=(A_{XY}^{I})^{\beta\beta\beta\alpha}$.

\begin{equation}
A_{LL}^{S}=-\frac{1}{2V^{4}m_A^{2}}\Big(V^{R^{\dagger}}B V^{L}\Big)^{\beta\beta}\Big(V^{R^{\dagger}}B V^{L}\Big)^{\beta\alpha}
\end{equation}
\begin{equation}
A_{LR}^{S}=\frac{1}{2V^{4}m_A^{2}}\Big(V^{R^{\dagger}}B V^{L}\Big)^{\beta\beta}\Big(V^{L^{\dagger}}B^\dagger V^{R}\Big)^{\beta\alpha}
\end{equation}\begin{equation}
A_{RL}^{S}=\frac{1}{2V^{4}m_A^{2}}\Big(V^{L^{\dagger}}B^\dagger V^{R}\Big)^{\beta\beta}\Big(V^{R^{\dagger}}B V^{L}\Big)^{\beta\alpha}
\end{equation}\begin{equation}
A_{RR}^{S}=-\frac{1}{2V^{4}m_A^{2}}\Big(V^{L^{\dagger}}B^\dagger V^{R}\Big)^{\beta\beta}\Big(V^{L^{\dagger}}B^\dagger V^{R}\Big)^{\beta\alpha}
\end{equation}

\subsection*{$\ell^{-}_{\alpha}\rightarrow \ell^{-}_{\beta}\ell^{-}_{\gamma}\ell^{+}_{\gamma}$}

There are two types of Feynman diagrams contributing to this
process. The first class involves a flavor conserving ($\gamma\gamma$)
and a flavor violating ($\beta\alpha$) vertex, while in the second
class both vertices violate flavor ($\beta\gamma$ and
$\gamma\alpha$). Therefore, the matching with Eq.~\eqref{eq:4lepton}
would yield non vanishing contributions to both coefficients
$(A_{XY}^{I})^{\gamma\gamma\beta\alpha}$ and
$(A_{XY}^{I})^{\beta\gamma\gamma\alpha}$. One can actually Fierz
transform the latter flavor structure into the former, thus in the
following expressions we set $A_{XY}^{I}=(
A_{XY}^{I})^{\gamma\gamma\beta\alpha}$. The Fierz transformations
involved in the matching are the following, where the type of
parenthesis indicates the fermion field which is contracted with the
gamma matrix in brackets.
\begin{equation}\begin{split}
(P_L)[P_L]=\frac{1}{2}(P_L][P_L)+\frac{1}{8}(\sigma^{\mu\nu}P_L][\sigma_{\mu\nu}P_L)\\
(P_R)[P_R]=\frac{1}{2}(P_R][P_R)+\frac{1}{8}(\sigma^{\mu\nu}P_R][\sigma_{\mu\nu}P_R)\\
(P_R)[P_L]=\frac{1}{2}(\gamma^{\mu}P_L][\gamma_{\mu}P_R)\\
(\gamma^{\mu}P_L)[\gamma_{\mu}P_L]=-(\gamma^{\mu}P_L][\gamma_{\mu}P_L)\\
(\gamma^{\mu}P_R)[\gamma_{\mu}P_R]=-(\gamma^{\mu}P_R][\gamma_{\mu}P_R)\\
(\gamma^{\mu}P_R)[\gamma_{\mu}P_L]=2(P_L][P_R)\\
\end{split}\end{equation}

\subsubsection*{$H_k$ contributions}

\begin{equation}
A_{LL}^{S}=\frac{1}{2m_{H_k}^{2}}\sum_{i,j}\Big[\big(W_{ki}c_i^{\gamma\gamma}\big)\big(W_{kj}c_j^{\beta\alpha}\big)-\frac{1}{2}\big(W_{ki}c_i^{\beta\gamma}\big)\big(W_{kj}c_j^{\gamma\alpha}\big)\Big]
\end{equation}

\begin{equation}
A_{LL}^{T}=\frac{1}{2m_{H_k}^{2}}\sum_{i,j}\Big[-\frac{1}{8}\big(W_{ki}c_i^{\beta\gamma}\big)\big(W_{kj}c_j^{\gamma\alpha}\big)\Big]
\end{equation}

\begin{equation}
A_{LR}^{S}=\frac{1}{2m_{H_k}^{2}}\sum_{i,j}\big(W_{ki}c_i^{\gamma\gamma}\big)\big(W_{kj} c_j^{\dagger^{\beta\alpha}}\big)
\end{equation}

\begin{equation}
A_{RL}^{S}=\frac{1}{2m_{H_k}^{2}}\sum_{i,j}\big(W_{ki}c_i^{\dagger^{\gamma\gamma}}\big)\big(W_{kj}c_j^{\beta\alpha}\big)
\end{equation}

\begin{equation}
A_{RR}^{S}=\frac{1}{2m_{H_k}^{2}}\sum_{i,j}\Big[\big(W_{ki}c_i^{\dagger{\gamma\gamma}}\big)\big(W_{kj}c_j^{\dagger{\beta\alpha}}\big)-\frac{1}{2}\big(W_{ki}c_i^{\dagger{\beta\gamma}}\big)\big(W_{kj}c_j^{\dagger{\gamma\alpha}}\big)\Big]
\end{equation}

\begin{equation}
A_{RR}^{T}=\frac{1}{2m_{H_k}^{2}}\sum_{i,j}\Big[-\frac{1}{8}\big(W_{ki}c_i^{\dagger{\beta\gamma}}\big)\big(W_{kj}c_j^{\dagger{\gamma\alpha}}\big)\Big]
\end{equation}

\begin{equation}
A_{LR}^{V}=\frac{1}{2m_{H_k}^{2}}\sum_{i,j}\Big[-\frac{1}{2}\big(W_{ki}c_i^{\beta\gamma}\big)\big(W_{kj}c_j^{\dagger{\gamma\alpha}}\big)\Big]
\end{equation}

\begin{equation}
A_{RL}^{V}=\frac{1}{2m_{H_k}^{2}}\sum_{i,j}\Big[-\frac{1}{2}\big(W_{ki}c_i^{\dagger{\beta\gamma}}\big)\big(W_{kj}c_j^{\gamma\alpha}\big)\Big]
\end{equation}

\subsubsection*{$Z$ contributions}

\begin{equation}
A_{LL}^{V}=-\frac{g^{2}}{4 m_W^{2}}\Big[\big(V^{L^{\dagger}}_{ee}V^{L}_{ee}-2\sin^{2}\theta_W\mathbb{I}\big)^{\gamma\gamma}\big(V^{L^{\dagger}}_{ee}V^{L}_{ee}\big)^{\beta\alpha}+\big(V^{L^{\dagger}}_{ee}V^{L}_{ee}\big)^{\beta\gamma}\big(V^{L^{\dagger}}_{ee}V^{L}_{ee}\big)^{\gamma\alpha}\Big]
\end{equation}
\begin{equation}
A_{LR}^{V}=0
\end{equation}\begin{equation}
A_{RL}^{V}=\frac{g^{2}}{2 m_W^{2}}\sin^{2}\theta_W\Big(V^{L^{\dagger}}_{ee}V^{L}_{ee}\Big)^{\beta\alpha}
\end{equation}\begin{equation}
A_{RR}^{V}=0
\end{equation}

The parametrization $V^{L(R)}=U^{L(R)}D^{L(R)}$ also simplifies the
expressions in this case. Thanks to the unitarity of $D_e^{L}$, we can
write
\begin{equation}\begin{split}
A_{LL}^{V}=\frac{g^{2}}{4 m_W^{2}M_F^{2}}\Big[\Big(1-2\sin^{2}\theta_W-\Big(D^{L^{\dagger}}_e \frac{m_S^{\dagger}m_S}{M_F^{2}}D^{L}_e\Big)^{\gamma\gamma}\Big)
\Big(D^{L^{\dagger}}_e m_S^{\dagger}m_S D^{L}_e\Big)^{\beta\alpha}\\
-\frac{1}{M_F^{2}}\Big(D^{L^{\dagger}}_e m_S^{\dagger}m_S D^{L}_e\Big)^{\beta\gamma}\Big(D^{L^{\dagger}}_e m_S^{\dagger}m_S D^{L}_e\Big)^{\gamma\alpha}\Big]
\end{split}\end{equation}

\begin{equation}
A_{RL}^{V}=-\frac{g^{2}}{2 m_W^{2}M_F^{2}}\sin^{2}\theta_W\Big(D^{L^{\dagger}}_e m_S^{\dagger}m_S D^{L}_e\Big)^{\beta\alpha}
\end{equation}

\subsubsection*{$A$ contributions}

\begin{equation}
A_{LL}^{S}=-\frac{1}{2V^{4}m_A^{2}}\Big[\big(V^{R^{\dagger}}B V^{L}\Big)^{\gamma\gamma}\Big(V^{R^{\dagger}}B V^{L}\big)^{\beta\alpha}-\frac{1}{2}\big(V^{R^{\dagger}}B V^{L}\Big)^{\beta\gamma}\Big(V^{R^{\dagger}}B V^{L}\big)^{\gamma\alpha}\Big]
\end{equation}

\begin{equation}
A_{LL}^{T}=-\frac{1}{2V^{4}m_A^{2}}\Big[-\frac{1}{8}\big(V^{R^{\dagger}}B V^{L}\Big)^{\beta\gamma}\Big(V^{R^{\dagger}}B V^{L}\big)^{\gamma\alpha}\Big]
\end{equation}

\begin{equation}
A_{LR}^{S}=\frac{1}{2V^{4}m_A^{2}}\Big(V^{R^{\dagger}}B V^{L}\Big)^{\gamma\gamma}\Big(V^{L^{\dagger}}B^\dagger V^{R}\Big)^{\beta\alpha}
\end{equation}

\begin{equation}
A_{RL}^{S}=\frac{1}{2V^{4}m_A^{2}}\Big(V^{L^{\dagger}}B^\dagger V^{R}\Big)^{\gamma\gamma}\Big(V^{R^{\dagger}}B V^{L}\Big)^{\beta\alpha}
\end{equation}

\begin{equation}
A_{RR}^{S}=-\frac{1}{2V^{4}m_A^{2}}\Big[\big(V^{L^{\dagger}}B^\dagger V^{R}\Big)^{\gamma\gamma}\Big(V^{L^{\dagger}}B^\dagger V^{R}\big)^{\beta\alpha}-\frac{1}{2}\big(V^{L^{\dagger}}B^\dagger V^{R}\Big)^{\beta\gamma}\Big(V^{L^{\dagger}}B^\dagger V^{R}\big)^{\gamma\alpha}\Big]
\end{equation}

\begin{equation}
A_{RR}^{T}=-\frac{1}{2V^{4}m_A^{2}}\Big[-\frac{1}{8}\big(V^{L^{\dagger}}B^\dagger V^{R}\Big)^{\beta\gamma}\Big(V^{L^{\dagger}}B^\dagger V^{R}\big)^{\gamma\alpha}\Big]
\end{equation}

\begin{equation}
A_{RL}^{V}=\frac{1}{2V^{4}m_A^{2}}\Big[-\frac{1}{2}\big(V^{L^{\dagger}}B^\dagger V^{R}\Big)^{\beta\gamma}\Big(V^{R^{\dagger}}B V^{L}\big)^{\gamma\alpha}\Big]
\end{equation}

\begin{equation}
A_{LR}^{V}=-\frac{1}{2V^{4}m_A^{2}}\Big[-\frac{1}{2}\big(V^{R^{\dagger}}B V^{L}\Big)^{\beta\gamma}\Big(V^{L^{\dagger}}B^\dagger V^{R}\big)^{\gamma\alpha}\Big]
\end{equation}

\subsection*{$\ell^{-}_{\alpha}\rightarrow \ell^{+}_{\beta}\ell^{-}_{\gamma}\ell^{-}_{\gamma}$}

In this process both vertices are necessarily flavor violating
($\gamma\beta$ and $\gamma\alpha$). This allows us to easily perform
the matching with Eq.~\eqref{eq:4lepton} and set in the following
expressions $A_{XY}^{I}=(A_{XY}^{I})^{\gamma\beta\gamma\alpha}$.

\subsubsection*{$H_k$ contributions}

\begin{equation}
A_{LL}^{S}=\frac{1}{2m_{H_k}^{2}}\sum_{i,j}\big(W_{ki}c_i^{\gamma\beta}\big)\big(W_{kj}c_j^{\gamma\alpha}\big)
\end{equation}

\begin{equation}
A_{LR}^{S}=\frac{1}{2m_{H_k}^{2}}\sum_{i,j}\big(W_{ki}c_i^{\gamma\beta}\big)\big(W_{kj} c_j^{\dagger^{\gamma\alpha}}\big)
\end{equation}

\begin{equation}
A_{RL}^{S}=\frac{1}{2m_{H_k}^{2}}\sum_{i,j}\big(W_{ki}c_i^{\dagger^{\gamma\beta}}\big)\big(W_{ki}c_j^{\gamma\alpha}\big)
\end{equation}

\begin{equation}
A_{RR}^{S}=\frac{1}{2m_{H_k}^{2}}\sum_{i,j}\big(W_{ki}c_i^{\dagger{\gamma\beta}}\big)\big(W_{kj}c_j^{\dagger{\gamma\alpha}}\big)
\end{equation}

\subsubsection*{$Z$ contributions}

\begin{equation} \label{eq:AVLLZ}
A_{LL}^{V}=-\frac{g^{2}}{4 m_W^{2}}\Big(V^{L^{\dagger}}_{ee}V^{L}_{ee}\Big)^{\gamma\beta}\Big(V^{L^{\dagger}}_{ee}V^{L}_{ee}\Big)^{\gamma\alpha}
\end{equation}
\begin{equation}
A_{LR}^{V}=0
\end{equation}\begin{equation}
A_{RL}^{V}=0
\end{equation}\begin{equation}
A_{RR}^{V}=0
\end{equation}

Finally, using our previous definitions we can simplify
Eq.~\eqref{eq:AVLLZ} to
\begin{equation}
A_{LL}^{V}=-\frac{g^{2}}{4 m_W^{2}M_F^{4}}\Big(D^{L^{\dagger}}_e m_S^{\dagger}m_S D^{L}_e\Big)^{\gamma\beta}\Big(D^{L^{\dagger}}_e m_S^{\dagger}m_S D^{L}_e\Big)^{\gamma\alpha}
\end{equation}

\subsubsection*{$A$ contributions}

\begin{equation}
A_{LL}^{S}=-\frac{1}{2V^{4}m_A^{2}}\Big(V^{R^{\dagger}}B V^{L}\Big)^{\gamma\beta}\Big(V^{R^{\dagger}}B V^{L}\Big)^{\gamma\alpha}
\end{equation}
\begin{equation}
A_{LR}^{S}=\frac{1}{2V^{4}m_A^{2}}\Big(V^{R^{\dagger}}B V^{L}\Big)^{\gamma\beta}\Big(V^{L^{\dagger}}B^\dagger V^{R}\Big)^{\gamma\alpha}
\end{equation}\begin{equation}
A_{RL}^{S}=\frac{1}{2V^{4}m_A^{2}}\Big(V^{L^{\dagger}}B^\dagger V^{R}\Big)^{\gamma\beta}\Big(V^{R^{\dagger}}B V^{L}\Big)^{\gamma\alpha}
\end{equation}\begin{equation}
A_{RR}^{S}=-\frac{1}{2V^{4}m_A^{2}}\Big(V^{L^{\dagger}}B^\dagger V^{R}\Big)^{\gamma\beta}\Big(V^{L^{\dagger}}B^\dagger V^{R}\Big)^{\gamma\alpha}
\end{equation}

%% file: tex/app3.tex
\section{$R_{h\mu\mu}$ analytical expression}
\label{sec:app3}

In our model, the $R_{h\mu\mu}$ ratio can be approximately written as
\begin{equation} \label{eq:RhmumuApp3}
  R_{h\mu\mu} = \frac{\text{BR}(h \to \mu\mu)}{\text{BR}(h \to \mu\mu)^{\rm SM}} \approx \left( \frac{c_{\mu\mu}^{S_H}+c_{\mu\mu}^{S_{\sigma}}+c_{\mu\mu}^{S_S}}{c_{\mu\mu}^{\rm SM}}\right)^2 \, ,
\end{equation}
where
\begin{equation}
c_{\mu\mu}^{\rm SM} = \frac{g m_{\mu}}{2 m_W}
\end{equation}
is the SM Higgs coupling to a pair of muons and $c_{\mu\mu}^{S_H}$,
$c_{\mu\mu}^{S_\sigma}$ and $c_{\mu\mu}^{S_S}$ denote the
contributions from the gauge eigenstates $S_H$, $S_\sigma$ and $S_S$,
respectively. These couplings are given by
\begin{align}
  c_{\mu\mu}^{S_H} &\approx\frac{1}{\sqrt{2}} Y_{e_{22}} V^{R^{\dagger}}_{22}V^{L^{\dagger}}_{22} \, \\
   c_{\mu\mu}^{S_\sigma} &\approx\frac{1}{\sqrt{2}} \sin \alpha\rho_{2}V^{R^{\dagger}}_{22}V^{L^{\dagger}}_{24} \, , \\
    c_{\mu\mu}^{S_S} &\approx\frac{1}{\sqrt{2}}\sin \beta Y_{S_2} V^{R^{\dagger}}_{24}V^{L^{\dagger}}_{22} \, , 
\end{align}
where we have assumed $\rho_1, \rho_3 \ll \rho_2$ and $Y_{S_1},
Y_{S_3} \ll Y_{S_2}$, as motivated by the explanation of the muon
$g-2$ anomaly and the stringent constraints from lepton flavor
violating observables. Furthermore, we have introduced the mixing
angles $\alpha$, $\beta$ and $\gamma$. The CP-even scalar mass matrix
$\mathcal{M}_R^2$ in Eq.\eqref{eq:MR2b} is diagonalized by the unitary
matrix $R$ which, assuming small mixing angles, can be parametrized as
\begin{equation} \label{eq:RotScalarmatrix}
  R = \left( \begin{array}{ccc}   
    1& \sin \alpha & \sin \beta \\
    -\sin \alpha&1&\sin \gamma\\
    -\sin \beta& -\sin \gamma &1 \end{array} \right) \, ,
\end{equation}
where $\alpha, \beta, \gamma \ll 1$. Using now
Eqs.~\eqref{eq.left}-\eqref{eq.light} we finally obtain
\begin{align} 
 c_{\mu\mu}^{S_H} &\approx \frac{m_{\mu}}{v_H}+\frac{Y_{S_{2}}v_S \rho_{2}v_{\sigma}}{2M_Fv_H} \left[ 1-\left(\frac{\rho_{2}v_{\sigma}}{2M_F}\right)^2 \right] \, , \label{eq:cmumu1} \\
 c_{\mu\mu}^{S_{\sigma}} &\approx \frac{\rho_{2}Y_{S_2}v_S}{2M_F} \, \sin\alpha \, , \label{eq:cmumu2} \\ 
 c_{\mu\mu}^{S_S} &\approx -\frac{\rho_{2}Y_{S_2}v_{\sigma}}{2M_F} \left[1-\left(\frac{\rho_{2}v_{\sigma}}{2M_F}\right)^2\right] \sin \beta \, . \label{eq:cmumu3}
\end{align}